\documentclass[onecolumn,a4paper,draft]{IEEEtran}
\IEEEoverridecommandlockouts
\usepackage{amsmath}
\usepackage{amssymb}
\usepackage{amsthm}
\usepackage{calc}
\usepackage{cite}
\usepackage{mathtools}
\usepackage{subcaption}
\usepackage{array,multirow}
\usepackage{arydshln}

\usepackage{tikz}
\usetikzlibrary{calc,decorations.pathreplacing,matrix,backgrounds, positioning}
\usepackage{pgfplots}
\tikzset{
	ncbar angle/.initial=90,
	ncbar/.style={
		to path=(\tikztostart)
		-- ($(\tikztostart)!#1!\pgfkeysvalueof{/tikz/ncbar angle}:(\tikztotarget)$)
		-- ($(\tikztotarget)!($(\tikztostart)!#1!\pgfkeysvalueof{/tikz/ncbar angle}:(\tikztotarget)$)!\pgfkeysvalueof{/tikz/ncbar angle}:(\tikztostart)$)
		-- (\tikztotarget)
	},
	ncbar/.default=0.5cm,
}

\tikzset{round left paren/.style={ncbar=0.5cm,out=100,in=-100}}
\tikzset{round right paren/.style={ncbar=0.5cm,out=80,in=-80}}

\tikzset{square left brace/.style={ncbar=0.1cm}}
\tikzset{square right brace/.style={ncbar=-0.1cm}}

\usepackage{color}
\definecolor{mygray}{gray}{0.6}

\newcolumntype{C}{>{$}c<{$}} 
\newcommand{\field}{\ensuremath{\mathbb{F}}}
\newcommand{\RS}{\ensuremath{\mathcal{RS}}}
\newcommand{\code}{\ensuremath{\mathcal{C}}}
\newcommand{\coded}{\ensuremath{\mathcal{D}}}

\newcommand{\wt}[1]{\ensuremath{\text{w}_\text{H} (#1)}}
\newcommand{\dcsd}{\ensuremath{d_{\code \star \coded}}}

\newcommand{\secref}[1]{Section~\ref{#1}}
\newcommand{\defref}[1]{Definition~\ref{#1}}
\newcommand{\lemref}[1]{Lemma~\ref{#1}}

\newcommand{\figref}[1]{Figure~\ref{#1}}

\newtheorem{theorem}{Theorem}

\newtheorem{lemma}{Lemma}
\newtheorem{corollary}{Corollary}
\newtheorem{definition}{Definition}
\newtheorem{example}{Example}
\newtheorem{remark}{Remark}

\DeclareMathOperator{\rk}{rk}
\DeclareMathOperator{\ord}{ord}

\definecolor{TUMBlau}{RGB}{0,101,189} 
\definecolor{TUMBlauDunkel}{RGB}{0,82,147} 
\definecolor{TUMBlauHell}{RGB}{152,198,234} 
\definecolor{TUMBlauMittel}{RGB}{100,160,200} 

\definecolor{TUMElfenbein}{RGB}{218,215,203} 
\definecolor{TUMGruen}{RGB}{162,173,0} 
\definecolor{TUMOrange}{RGB}{227,114,34} 
\definecolor{TUMGrau}{gray}{0.6} 

\definecolor{TUMGruenHell}{RGB}{0,124,48}
\definecolor{TUMRot}{RGB}{196,7,27}

\usepackage{letltxmacro}
\LetLtxMacro\orgvdots\vdots
\LetLtxMacro\orgddots\ddots

\makeatletter
\DeclareRobustCommand\vdots{%
  \mathpalette\@vdots{}%
}
\newcommand*{\@vdots}[2]{%
  \sbox0{$#1\cdotp\cdotp\cdotp\m@th$}%
  \sbox2{$#1.\m@th$}%
  \vbox{%
    \dimen@=\wd0 %
    \advance\dimen@ -3\ht2 %
    \kern.5\dimen@
    \dimen@=\wd2 %
    \advance\dimen@ -\ht2 %
    \dimen2=\wd0 %
    \advance\dimen2 -\dimen@
    \vbox to \dimen2{%
      \offinterlineskip
      \copy2 \vfill\copy2 \vfill\copy2 %
    }%
  }%
}
\DeclareRobustCommand\ddots{%
  \mathinner{%
    \mathpalette\@ddots{}%
    \mkern\thinmuskip
  }%
}
\newcommand*{\@ddots}[2]{%
  \sbox0{$#1\cdotp\cdotp\cdotp\m@th$}%
  \sbox2{$#1.\m@th$}%
  \vbox{%
    \dimen@=\wd0 %
    \advance\dimen@ -3\ht2 %
    \kern.5\dimen@
    \dimen@=\wd2 %
    \advance\dimen@ -\ht2 %
    \dimen2=\wd0 %
    \advance\dimen2 -\dimen@
    \vbox to \dimen2{%
      \offinterlineskip
      \hbox{$#1\mathpunct{.}\m@th$}%
      \vfill
      \hbox{$#1\mathpunct{\kern\wd2}\mathpunct{.}\m@th$}%
      \vfill
      \hbox{$#1\mathpunct{\kern\wd2}\mathpunct{\kern\wd2}\mathpunct{.}\m@th$}%
    }%
  }%
}
\makeatother

\begin{document}

\title{Private Streaming with Convolutional Codes}

 \author{Lukas Holzbaur, Ragnar Freij-Hollanti, Antonia Wachter-Zeh, Camilla Hollanti \thanks{The work of L. Holzbaur, R. Freij-Hollanti, and A. Wachter-Zeh was supported by the Technical University of Munich -- Institute for Advanced Study, funded by the German Excellence Initiative and European Union 7th Framework Programme under Grant Agreement No. 291763 and the German Research Foundation (Deutsche Forschungsgemeinschaft, DFG) under Grant No. WA3907/1-1.
     The work of C. Hollanti was supported by the Academy of Finland, under Grants No. 276031, 282938, and 303819, and by the Technical University of Munich -- Institute for Advanced Study, funded by the German Excellence Initiative and the EU 7th Framework Programme under Grant Agreement No. 291763, via a Hans Fischer Fellowship.}%
   \thanks{Parts of this paper have been presented at the \emph{2018 IEEE Information Theory Workshop (ITW 2018)} \cite{Holzbaur2018}.}%
\thanks{L. Holzbaur and A. Wachter-Zeh are with the Institute for Communications Engineering, Technical University of Munich, Germany. R.~Freij-Hollanti and C. Hollanti are with the Department of Mathematics and Systems Analysis, Aalto University, Finland. }}

\maketitle

\begin{abstract}
  Recently, information-theoretic private information retrieval (PIR) from coded storage systems has gained a lot of attention, and a general star product PIR scheme was proposed. In this paper, the star product scheme is adopted, with appropriate modifications, to the case of private (\emph{e.g.}, video) streaming. It is assumed that the files to be streamed are stored on~$n$ servers in a coded form, and the streaming is carried out via a convolutional code. The star product scheme is defined for this special case, and various properties are analyzed for two channel models related to straggling and Byzantine servers, both in the baseline case as well as with colluding servers. The achieved PIR rates for the given models are derived and, for the cases where the capacity is known, the first model is shown to be asymptotically optimal, when the number of stripes in a file is large.  The second scheme introduced in this work is shown to be the equivalent of block convolutional codes in the PIR setting. For the Byzantine server model, it is shown to outperform the trivial scheme of downloading stripes of the desired file separately without memory.
\end{abstract}

\IEEEpeerreviewmaketitle

\section{Introduction}

Private information retrieval (PIR) studies the problem when a user wants to retrieve a file from a storage system without revealing the identity of the file in question to the storage servers. The original problem was introduced in \cite{chor1995private,PIR_original}, and more recently the problem setting was extended to the case where the files are stored on the servers in an encoded form rather than merely being replicated \cite{shah2014,fazeli2015pir,razan_salim}. The capacity of PIR for replicated storage was derived in \cite{sun_jafar_1} and for coded storage in \cite{Banawan2018}. For the case of colluding servers, \emph{i.e.}, servers that cooperate to determine the index of the requested file, the capacity was derived in \cite{Sun2016} and for colluding and Byzantine servers in \cite{Banawan2017}. In \cite{Wang2017,Wang20172,Holzbaur2019} the respective capacities of \emph{symmetric} PIR were derived. In \cite{freij2016private}, a so-called star product PIR scheme was introduced. The scheme works with any linear code as a storage code and retrieval code, and the highest rate possible for this scheme is achieved when both codes are generalized Reed-Solomon (GRS) codes.

Currently, Netflix and Youtube alone are occupying more than 50\% of Internet downstream traffic. Motivated by this huge increase in multi-media streaming,  we will consider \emph{private streaming} suitable for distributed systems sharing encoded streams. In a wider context, this is related to the problem of private stream search (PSS), which has been considered, \emph{e.g.}, in \cite{search-stream2005,Bethencourt2009NewTF, Finiasz2012PrivateSS}, typically using cryptographic assumptions, and allows the user to privately learn the contents of the servers. As in most works on information-theoretic PIR, we assume the user knows these contents and is able to query each server for linear combinations of files. In this paper, we require information-theoretic privacy, namely that the servers gain zero information on the index of the file being requested for streaming, based on the query received from the user.

Streaming applications require low latency decoding of the received data blocks and it has been shown that, under such constraints, convolutional codes perform well \cite{Badr2013,Kuijper2016} for different erasure channels. When considering errors, convolutional codes are sensitive to burst errors but good at handling well-distributed errors \cite{Dettmar1995}. As burst errors are unlikely on, \emph{e.g.}, an additive white Gaussian noise (AWGN) channel, they exhibit good performance compared to block codes on such channels and have a lower bit error rate than comparable block codes with the same rate~\cite[Section V]{Dettmar1995}. Compared to other codes that have also been shown to be well suited for streaming, \emph{e.g.}, fountain codes \cite{Vukobratovic2009}, a further advantage of convolutional codes is that they have a partial block structure. This allows for a natural combination of convolutional codes with the star product scheme, which is an efficient and flexible  PIR scheme\cite{tajeddine2017private,tajeddine2018byzantine}. 

By the use of convolutional codes, the presented scheme is related to PIR from databases encoded with non-MDS codes, for which constructions achieving the MDS PIR capacity without collusion~\cite{Banawan2018} exist~\cite{Freij-Hollanti20172}~\cite{Kumar2017}.

The main contributions of this paper are the following.
\begin{itemize}
\item To the best of the authors' knowledge, information-theoretically private streaming is considered for the first time.
\item Memory is introduced into the star product PIR scheme by a block convolutional structure, improving the performance of the decoder for a large class of channels. 
\item Two schemes for different channels, namely a block erasure channel and a non-bursty channel, \emph{e.g.}, an AWGN channel, are given. Both can operate on the same database and the user can adapt the queries according to the current channel conditions.
\item The achieved PIR rates are derived and for the block erasure scheme shown to be either asymptotically optimal for the considered model, or, for cases where the capacity is unknown, shown to coincide with conjectures on the asymptotic capacity. For the Byzantine server model, the introduced scheme is shown to outperform the straight forward scheme of downloading stripes of the desired file separately without memory.
\end{itemize}
This paper is structured as follows. In Section~\ref{sec:preliminaries}, we briefly introduce block convolutional codes and the star product PIR scheme of~\cite{freij2016private}. In Section~\ref{sec:pirconv}, we describe the combination of block convolutional codes with the star product PIR scheme and show that the achievable asymptotic (in the number of stripes and files) PIR rate is equal to the conjectured PIR capacity of a coded scheme with $t$-collusion. In Section~\ref{sec:block}, the scheme is adapted for a block erasure channel in which the user does not receive replies in a given number of consecutive iterations of the protocol. In Section~\ref{sec:byzantine}, we introduce a scheme for non-bursty channels based on the decoding algorithm of~\cite{Dettmar1995}.

\section{Preliminaries}\label{sec:preliminaries}

We denote by~$[a,b]$ the set of integers~$\{ i \; | \; a\leq i \leq b \}$ and~$[b]=[1,b]$. Throughout the paper,~$\field$ will denote an arbitrary finite field, $\left\langle \cdot \right\rangle$ denotes the linear span, and $\left\langle \cdot, \cdot \right\rangle$ denotes the inner product.

If~$c$ and~$d$ are vectors of the same length~$n$, we define their \emph{star product} as the coordinate-wise product
\begin{equation*}
c\star d = \left(c_1 d_1, \dots , c_n d_n\right) \ .
\end{equation*}
Further, if~$\code$ and~$\coded$ are linear codes of the same length, we define their star product to be the linear code given by the span of the pairwise star product of codewords from $\code$ and $\coded$, \emph{i.e.},
\begin{equation*}
    \code\star\coded = \left\langle c\star d \,|\, c\in\code, d\in \coded\right\rangle \ .
\end{equation*}

\subsection{Convolutional Codes}

\begin{definition}[Convolutional code]\label{def:convcode}
  Let~$G_1,\ldots, G_{M+1} \in \field^{k\times n}$ and~$\mathrm{rank} (G_1) = k$. Define an~$(n,k)$ \emph{convolutional code}~$\code_c$ as
  \begin{equation}\label{eq:convcode}
    Y_i = \sum_{j=1}^{M+1} X_{i-j+1} G_{j} \ ,
  \end{equation}
  where~$X_{0} ,\ldots, X_{-M+1} = 0$ and~$X_j \in \field^{k}$.
\end{definition}
We refer to~$M$ as the \emph{memory} of~$\code_c$, and if~$M=1$, we say that~$\code_c$ is a \emph{unit memory} (UM) code. In this paper, we consider \emph{terminated} convolutional codes, \emph{i.e.},~$Y$ is not a semi-infinite vector, but~$Y = (Y_1,Y_2,\dots,Y_{\ell+M})$ where~$Y_i$ is defined as in~\eqref{eq:convcode}.

An~$(n,k)$-code denotes a linear block code of length~$n$ and dimension~$k$.
A generalized  Reed--Solomon (GRS) code~$\RS(n,k,v)$ is an~$(n,k)$-code with minimum distance~$d = n-k+1$ and generator matrix
\begin{align*}
G = \left(\begin{matrix}
1&1^{\vphantom{k-1}}&\cdots&1 \\
\alpha_1&\alpha_2^{\vphantom{k-1}}&\cdots&\alpha_{n} \\
\vdots&\vdots&\ddots&\vdots \\
\alpha_1^{k-1}&\alpha_2^{k-1}&\cdots&\alpha_{n}^{k-1} \\
\end{matrix} \right)
\left(\begin{matrix}
v_1^{\vphantom{k-1}}&0&\cdots&0\\
0&v_2^{\vphantom{k-1}}&\cdots&0\\
\vdots&\vdots&\ddots&\vdots\\
0&0&\cdots&v_n^{\vphantom{k-1}}
\end{matrix}\right) \ ,
\end{align*}
where~$\alpha_1,\dots \alpha_n \in \field$ are distinct \emph{evaluation points} and the~$v_j$'s are all non-zero. If the choice of the~$v_j$'s is not important, we sometimes write~$\RS(n,k)$ code. It is well known that $\RS$ codes are maximum distance separable (MDS), i.e., fulfill the Singleton bound
\begin{equation}
  d \leq n-k+1
  \label{eq:singletonBound}
\end{equation}
with equality (see, e.g., \cite{MS77}).

The distance measure of interest for convolutional codes is the \emph{extended row distance}~$d_\iota^r$, which determines the minimum number of errors required for an error burst of~$\iota$ blocks to occur. For UM codes this distance can be lower bounded by the \emph{designed extended row distance} \cite{Dettmar1995} 
\begin{equation}
\label{eq:designedextrowdist}
\bar{d}_\iota^r = d_1 + (\iota-1)d_\alpha + d_2, \; \iota \geq 1 \ ,
\end{equation}
where~$d_1$,~$d_2$, and~$d_\alpha$ denote the distances of the codes generated by~$G_1$,~$G_2$, and~$(G_1^T,G_2^T)^T$ respectively. In \cite{Dettmar1995}, a decoding algorithm is given, which is guaranteed to be successful if the number of errors does not exceed half the designed extended row distance for any~$\iota$, \emph{i.e.}, it decodes successfully if
\begin{equation}
\label{eq:decodingsuccess}
\sum_{j=s}^{\iota+s} \wt{w_j} < \frac{\bar{d}_\iota^r}{2}, \; \forall \; s \in [\ell+M], \iota \in [0,\ell+M-s] \ ,
\end{equation}
where~$w_j$ denotes the error vector of the~$j$-th block.
We refer to an~$(n,k)$ UM code for which~$d_\alpha$,~$d_1$ and~$d_2$ fulfill the Singleton bound for block codes, as given in (\ref{eq:singletonBound}), with equality as an \emph{optimal}~$(n,k)$ UM code.

\subsection{Star Product PIR}\label{StarProd}
We review the star product scheme for PIR from an arbitrary storage code, as introduced in~\cite{freij2016private}. Let~$\code$ be an~$(n,k)$ code (the \emph{storage} code) with generator matrix $G\in\field^{k\times n}$, storing~$m$ independent files~$X^1,\ldots , X^m \in \field^k$. Each file is drawn i.i.d. randomly from $\field$, thereby 
\begin{align*}
  H(X^s) = k \log(|\field|), \ \forall \ s \in [m] \\
  H(X^1,...,X^m) = mk \log(|\field|) \ ,
\end{align*}
where $|\field|$ denotes the order of $\field$.
Note, that for simplicity we do not consider subpacketization/striping of the files here, for more details see \cite{freij2016private}.
Each server~$j\in [n]$ stores a column~$Y_{j}$ of the matrix~$Y=XG\in \field^{m\times n}$, where~$X\in \field^{m\times k}$ is a \emph{data matrix}, whose~$i$-th row~$X^i$ represents the~$i$-th file. The scheme we will describe allows a user to retrieve the file~$X^i$ without disclosing the index~$i$.

Let~$\coded$ be a code of the same length~$n$ as~$\code$. Let~$D \in \field^{m\times n}$ be a matrix whose~$m$ rows are i.i.d. uniformly random codewords of~$\coded$.
The query for the~$j$-th server is given by
\begin{equation}
  \label{eq:query}
  q^i_j = D_{\cdot,j} + e_i E_{1,j} \ ,
\end{equation}
where~$e_i$ denotes the~$i$-th standard basis vector and\footnote{This notation is chosen to be consistent with the
  later sections when~$E$ will be a matrix.}~$E=E_{1,\cdot} \in \field^{1\times n}$, where the $\cdot$ as index means that we consider all columns.

The servers now respond with the standard inner product of their~$(m\times 1)$ stored vector~$Y_{j}$ and the query
vector~$q_j^i$ which they received, so the response of the~$j$-th server is the symbol
\begin{equation}\label{eq:response2}
  r_{j}^i = \left\langle q_j^i , Y_{j} \right\rangle = \sum_{s=1}^{m} D_{s,j} Y_{j}^s + E_{1,j} Y_{j}^i \in \field \ .
\end{equation}

Considering the~$n$ responses obtained as a vector in~$\field^{1\times n}$, we can write it as
\begin{align*}
  r^i= & \sum_{s=1}^{m} \left(D_{s,1} Y_{1}^s, \dots , D_{s,n} Y_{n}^s \right) + \left(E_{1,1} Y_{1}^i, \dots , E_{1,n} Y_{n}^i\right) \; \in  \code\star\coded + E \star Y_{}^i \ .
\end{align*}
Assuming~$E$ has weight~$\wt E < d_{\code\star\coded}$\,, where $d_{\code\star\coded}$ is the distance of $\code \star \coded$, erasure decoding in~$\code\star \coded$ now allows us to retrieve the vector~$E \star Y^i$, which depends only on the desired file~$Y^i$.
The rate achievable by this scheme is 
\begin{equation}
  \label{eq:RPIRorig}
  R^{\star}_{\mathrm{PIR}} = \frac{n-(k+t-1)}{n} \ ,
\end{equation}
where~$t$ is the number of colluding servers, \emph{i.e.}, the maximal number of servers that can exchange their queries such that the scheme is still private (see \cite{freij2016private} for details). If the response is corrupted by channel erasures or a bounded number of malicious servers, the user first decodes the response in $\code\star(\coded+E)$, as in~\cite{tajeddine2018byzantine}. This is discussed further in Section~\ref{sec:byzantine}. Generally the star product scheme works with any storage and retrieval code, the rate however depends on the distance of the star product of these codes. One class of codes that achieves the highest possible rate this scheme can achieve, is the class of GRS codes, where storage and retrieval code have the same code locators.

\section{PIR from Convolutional Codes} \label{sec:pirconv}

In this section, it is shown how a large file can be streamed with asymptotically (with respect to the number of stripes in each file) no rate loss compared to the retrieval of stripes without memory, by designing, as per user's request, the retrieved symbols such that they are codewords of a block convolutional code of memory~$M$. By itself, this does not offer any advantage compared to the star product scheme, but it gives the basis of the improvements presented in Section~\ref{sec:block} and~\ref{sec:byzantine}.

In the star product scheme, as introduced in Section~\ref{StarProd}, the user is able to retrieve $d_{\code \star \coded}-1$ symbols of the encoded desired file. It is easy to verify that instead of only adding one vector $E_{1,\cdot}$ in the $i$-th row in~(\ref{eq:query}), the user can also add multiple vectors $E_{1,\cdot}, E_{2,\cdot},...$ of the same support in multiple rows. This allows for the recovery of linear combinations of encoded symbols of the files corresponding to these rows, where the coefficients of the linear combinations are given by the entries of the $E$ vectors. Trivially, if each file consists of multiple stripes, linear combinations of these stripes can be retrieved. We use this inherent property of the star product scheme to design the queries such that these linear combinations have a special structure resembling a block convolution code.

Note that for the scheme introduced in this section the requirements on the storage and retrieval code are only given by the star product scheme, i.e., their star product should be of large distance. As will be discussed further in Section~\ref{sec:block} and Section~\ref{sec:byzantine}, the schemes introduced there have additional requirements on the codes. Since GRS codes fulfill these requirements and are well-suited for block convolutional codes, the star product scheme, and distributed data storage in general, we focus on this class of codes.

\subsection{Storage Code}

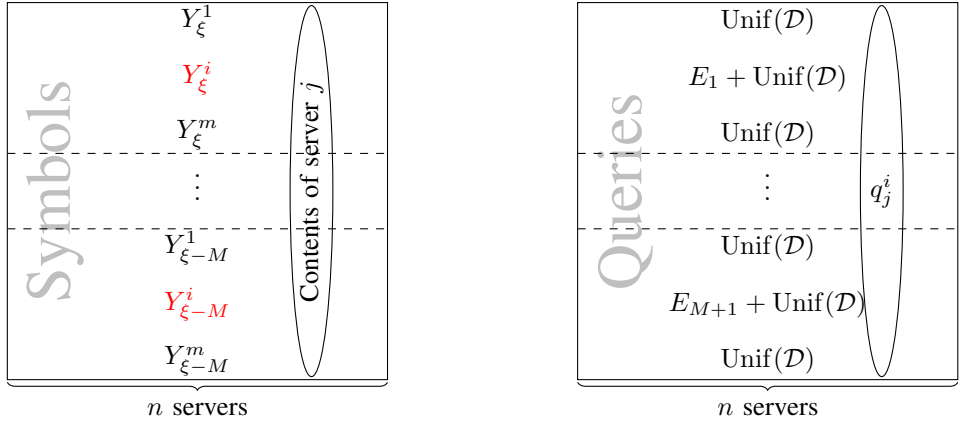
\begin{figure}
\centering
\def\x{*0.5}

\begin{tikzpicture}
	
\node [draw = none,rotate = 90] (symbols) at (1.25\x,5\x) {\color{lightgray} \Huge Symbols};
\draw (0\x,0\x) rectangle (10\x,10\x);
\draw [decorate,decoration={brace,amplitude=3pt},xshift=0pt,yshift=-1pt]
(10\x,0\x) -- (0\x,0\x) node [black,midway, yshift=-10pt] 
{$n$ servers};
\draw [dashed] (0\x,4\x) -- (10\x,4\x);
\node [draw = none] (vdots) at (5\x,5.25\x) {$\vdots$};
\draw [dashed] (0\x,6\x) -- (10\x,6\x);
\node [draw = none] (yMm) at (5\x,0.5\x) {$Y^m_{\xi-M}$};
\node [draw = none, color = red] (yMi) at (5\x,2\x) {$Y^i_{\xi-M}$};
\node [draw = none] (yM1) at (5\x,3.5\x) {$Y^1_{\xi-M}$};
\node [draw = none] (y1m) at (5\x,6.5\x) {$Y^m_{\xi}$};
\node [draw = none, color = red] (y1i) at (5\x,8\x) {$Y^i_{\xi}$};
\node [draw = none] (y11) at (5\x,9.5\x) {$Y^1_{\xi}$};
\draw (8\x,5\x) ellipse (8pt and 70pt);
\node [rotate=90] (content) at (8\x,5\x) { Contents of server $j$ };

\node [draw = none,rotate = 90] (queries) at (16.25\x,5\x) {\color{lightgray} \Huge Queries};
\draw (15\x,0\x) rectangle (25\x,10\x);
\draw [decorate,decoration={brace,amplitude=3pt},xshift=0pt,yshift=-1pt]
(25\x,0\x) -- (15\x,0\x) node [black,midway, yshift=-10pt] 
{$n$ servers};
\draw [dashed] (15\x,4\x) -- (25\x,4\x);
\node [draw = none] (vdots) at (20\x,5.25\x) {$\vdots$};
\draw [dashed] (15\x,6\x) -- (25\x,6\x);
\node (qMm) at (20\x,0.5\x) {$\rm{Unif}(\coded)$};
\node (qMi) at (20\x,2\x) {$E_{M+1} + \rm{Unif}(\coded)$};
\node (qM1) at (20\x,3.5\x) {$\rm{Unif}(\coded)$};
\node (q1m) at (20\x,6.5\x) {$\rm{Unif}(\coded)$};
\node (q1i) at (20\x,8\x) {$E_1 + \rm{Unif}(\coded)$};
\node (q11) at (20\x,9.5\x) {$\rm{Unif}(\coded)$};
\draw (23\x,5\x) ellipse (8pt and 70pt); 
\node (ellipse) at (23\x,5\x) {$q^i_j$};

\end{tikzpicture}
\caption{The queried symbols in iteration~$\xi$ and the query matrix, where~$\rm{Unif}(\coded)$ denotes i.i.d. uniformly random codewords from~$\coded$. The~$j$-th server responds with the inner product of the two vectors marked with ellipses.}
\label{fig:queries}
\end{figure}

Denote by~$m$ the number of files~$X^1, ..., X^m \in \field^{\ell k}$ and by~$n$ the number of servers. Each file is drawn i.i.d. randomly from $\field$, thereby 
\begin{align*}
  H(X^s) = \ell k \log(|\field|), \ \forall \ s \in [m] \\
  H(X^1,...,X^m) = m\ell k \log(|\field|) \ ,
\end{align*}
where $|\field|$ denotes the order of $\field$.
 The files are split into~$\ell$ stripes~$X^s_i \in \field^k$ and encoded with an~$\RS(n,k)$ storage code~$\code$ with evaluation points~$\alpha_j,\ j\in [n]$. The~$j$-th server stores the~$j$-th symbol of every encoded stripe~$Y^s_i \in \field^n$ (see \figref{fig:queries}).

\subsection{Query} \label{subsec:queryconv}

The query is designed such that the (encoded) symbols of the desired file retrieved from the servers responses form a convolutional code of memory $M$, with $0\leq M \leq \ell-1$. To achieve this, the user queries for carefully chosen linear combinations of~$M+1$ stripes in each block.
Let~$\coded$ be an~$\RS(n,t)$ code; the matrix~$D\in \field^{(M+1)m\times n}$ as in \eqref{eq:query}; and~$J\subset [n]$ with~$|J| \leq d_{\code \star \coded} -1$. The query for the~$j$-th server is given by\footnote{Note that~$E_{z+1,j}$ is a scalar.}
\begin{equation}
  \label{eq:queryconv}
  q_j^i = D_{\cdot,j} + e_{zm+i} E_{z+1,j} , \; z \in [0,M] \ ,
\end{equation}
where~$e_i$ is the~$i$-th standard basis vector and the matrix~$E \in \field^{M+1\times n}$ is given by
\begin{equation}
  \label{eq:Evector}
  E_{z+1,j} = \left\{
    \begin{array}{ll}
      \alpha_j^{zk}, & \mathrm{\ if}\; j\in J \\
      0, & \mathrm{\ otherwise}
    \end{array} \right. \ .
\end{equation}

\subsection{Response} \label{subsec:responseconv}

The protocol consists of~$\ell+M$ iterations in each of which the servers respond with the inner product of the query and a vector containing the stored symbols of~$M+1$ stripes of each file, depending on the iteration. In iteration~$\xi$ the response of server~$j$ is given by
\begin{align}
  \label{eq:responseconvj}
  r_{\xi,j}^i &= \left\langle q_j^i , (Y_{\xi , j},Y_{\xi-1, j},\ldots,Y_{\xi-M, j})^T \right\rangle \ ,
\end{align}
where~$Y_{-M+1} = \cdots = Y_0 = Y_{\ell+1} = \cdots = Y_{\ell +M} = 0$ and~$Y_{\xi}=X_{\xi}G$ denotes the matrix storing the~$\xi$-th part of every file.

\subsection{Decoding} \label{subsec:decodingconv}
The response is given by
\begin{equation}
  \label{eq:responseconv}
r_{\xi}^i = \sum_{z=0}^{M} \underbrace{\sum_{s=1}^m Y_{\xi-z}^s \star D_{zm+s}}_{\in \code \star \coded} + \underbrace{Y_{\xi-z}^i \star E_{z+1}}_{\in \code \star E_{z+1}} \ .
\end{equation}
An illustration of the responses for the case of $M=1$ is given in \figref{fig:responses}. 
\begin{lemma}
Let~$|J| \geq k$. Given the the responses~$\{r_{1}^i, r_{2}^i, \ldots, r_{\ell}^i\}$ the file~$X^i$ can be recovered.
\end{lemma}
\begin{IEEEproof}
By (\ref{eq:Evector}) the vectors~$E_{z+1}$ are designed such that for any~${c \in \code \star E_{z+1}}$,~$z \in [0,M]$ it holds that~$c_j = 0$,~$\forall \, j \notin J$. As~$|J| \leq d_{\code \star \coded}-1$ erasure decoding in~$\code \star \coded$ recovers the vector 
\begin{equation*}
\sum_{z=0}^{M} E_{z+1} \star Y_{\xi-z}^i = \sum_{z=0}^{M} X_{\xi-z}^i \cdot G_{\code \star E_{z+1}}
\end{equation*}
in each iteration, where the~$G_{\code \star E_{z+1}}$ are generator matrices of the storage code~$\code$ with column multipliers~$E_{z+1}$. Since~$|J| \geq k$, each~$G_{\code \star E_{z+1}}$ is of rank~$k$ and it follows that given the set~$\{ X_{\xi-M}^i,\ldots,X_{\xi}^i\} \backslash X_z^i$, the stripe~$X_z^i$ can be determined uniquely. In the first iteration~$X_{1-M} = \cdots = X_{0} = 0$ so~$X_{1}$ can be recovered and recovery of the remaining stripes follows by induction.
\end{IEEEproof}

As both~$\code$ and~$\coded$ are GRS codes, the distance of the star product~$\code \star \coded$ is given by~$\dcsd = n- (k+t-1)+1$ and it follows that at most~$\dcsd -1= n-(k+t-1)$ symbols can be downloaded in each iteration. In order to retrieve each stripe in one iteration, we require $k\leq n-(k+t-1)$. As there are~$\ell+M$ iterations the total number of downloaded symbols of $\field$ is $(\ell+M)n$. Hence, the PIR rate achievable by this scheme is
\begin{align}
  \label{eq:pirrate}
  R_{\mathrm{PIR}} &= \frac{H(X^i)}{(\ell+M)n \log(|\field|)}\nonumber\\
                   &= \frac{\ell k \log(|\field|)}{(\ell+M)n \log(|\field|)}\nonumber\\ 
                   &\leq \frac{\ell (n-(k+t-1)) }{(\ell+M)n} \nonumber\\
                   &= \frac{\ell (n-(k+t-1))}{(\ell+M) n} \ . 
\end{align}
Some observations are in order:
\begin{enumerate}
\item For any given memory $M$, the upper bound on the rate approaches the PIR rate of \cite{freij2016private} given in (\ref{eq:RPIRorig}) for~$\ell \rightarrow \infty$.
\item The highest PIR rate in this setting is achieved for~$|J| = k$ and~$n = 2k+t-1$.
\item For the trivial case of memory $M=0$ the scheme is a repeated application of the star product scheme \cite{freij2016private} and therefore asymptotically, \emph{i.e.,} for $m \rightarrow \infty$, achieves the capacity of PIR from MDS coded databases without collusion \cite{Banawan2018} and the recently proved capacity of linear PIR from MDS coded databases for $t\geq 1$ under some restrictions~\cite{Holzbaur2019capacity}. 
\end{enumerate}

\begin{figure}
  \centering
  \def\x{1}

\begin{tikzpicture}

\coordinate (d0) at (\x*1,\x*5);
\coordinate (d1) at (\x*2,\x*4.5);
\coordinate (d2) at (\x*3,\x*4);
\coordinate (d3) at (\x*4,\x*3.5);
\coordinate (d4) at (\x*5,\x*3);
\coordinate (d5) at (\x*6,\x*2.5);
\coordinate (d6) at (\x*7,\x*2);

\draw (d0) rectangle (d1);
\draw (d1) rectangle (d2);
\draw (d2) rectangle (d3);
\draw (d4) rectangle (d5);
\draw (d5) rectangle (d6);

\node[draw=none] (R1) at ($(d0) + (\x*0.5,-\x*0.25)$) {$G_{\code \star E_1}$};
\node[draw=none] (R1) at ($(d1) + (\x*0.5,-\x*0.25)$){$G_{\code \star E_1}$};
\node[draw=none] (R1) at ($(d2) + (\x*0.5,-\x*0.25)$){$G_{\code \star E_1}$};
\node[draw=none] (R1) at ($(d3) + (\x*0.5,-\x*0.2)$){$\ddots$};
\node[draw=none] (R1) at ($(d4) + (\x*0.5,-\x*0.25)$){$G_{\code \star E_1}$};
\node[draw=none] (R1) at ($(d5) + (\x*0.5,-\x*0.25)$){$G_{\code \star E_1}$};

\node[draw=none] (csd) at ($(d0) + (-\x*4.2,-\x*1.5)$) {$\code \star \coded \quad +$};


\coordinate (dd0) at (\x*2,\x*5);
\coordinate (dd1) at (\x*3,\x*4.5);
\coordinate (dd2) at (\x*4,\x*4);
\coordinate (dd3) at (\x*5,\x*3.5);
\coordinate (dd4) at (\x*6,\x*3);
\coordinate (dd5) at (\x*7,\x*2.5);
\coordinate (dd6) at (\x*8,\x*2);

\draw (dd0) rectangle (dd1);
\draw (dd1) rectangle (dd2);
\draw (dd2) rectangle (dd3);
\draw (dd4) rectangle (dd5);
\draw (dd5) rectangle (dd6);

\node[draw=none] (R1) at ($(dd0) + (\x*0.5,-\x*0.25)$) {$G_{\code \star E_2}$};
\node[draw=none] (R1) at ($(dd1) + (\x*0.5,-\x*0.25)$){$G_{\code \star E_2}$};
\node[draw=none] (R1) at ($(dd2) + (\x*0.5,-\x*0.25)$) {$G_{\code \star E_2}$};
\node[draw=none] (R1) at ($(dd3) + (\x*0.5,-\x*0.2)$) {$\ddots$};
\node[draw=none] (R1) at ($(dd4) + (\x*0.5,-\x*0.25)$) {$G_{\code \star E_2}$};
\node[draw=none] (R1) at ($(dd5) + (\x*0.5,-\x*0.25)$) {$G_{\code \star E_2}$};


\draw [] ($(d0) + (-\x*0.2,-\x*3)$) to [round left paren ] ($(d0) + (-\x*0.2,-\x*0)$);
\draw [] ($(d6) + (\x*1.2,-\x*0)$) to [round right paren] ($(d6) + (\x*1.2,\x*3)$);


\draw [decorate,decoration={brace,amplitude=3pt},xshift=0pt,yshift=0pt]
($(d1) + (-\x*0.05,-\x*0.5)$) -- ($(d1) + (-\x*0.05,\x*0)$) node [black,midway,xshift=-\x*0.4cm] {$k$};

\draw [decorate,decoration={brace,amplitude=3pt},xshift=0pt,yshift=0pt]
($(d2) + (\x*0,-\x*0.05)$) -- ($(d2) + (-\x*1,-\x*0.05)$) node [black,midway,yshift=-\x*0.4cm] {$n$};

\draw [dashed, color=red] (d0) -- ($(d0) + (\x*0,-\x*3.5)$);
\draw [dashed, color=red] ($(d3) + (\x*0,\x*1.5)$) -- ($(d3) + (\x*0,-\x*2)$);

\node[draw = none, color=red] (tw) at ($(d0) + (\x*1.5,-\x*3.75)$) {decoding window};

\draw [->, color=red] ($(d0) + (\x*1,-\x*3.25)$) -- ($(d0) + (\x*0,-\x*3.25)$);
\draw [->, color=red] ($(d3) + (-\x*1,-\x*1.75)$) -- ($(d3) + (\x*0,-\x*1.75)$);
\node[draw = none,color=red] at ($(d0) + (\x*1.5,-\x*3.25)$) {$N$};


\node (mult) at ($(d0) + (-\x*0.625,-\x*1.5)$) {$\cdot$};

\coordinate (xl) at ($(d0) + (-\x*1.25,-\x*1.5)$);

\draw [] ($(xl) + (\x*0.25,-\x*0.3)$) to [square right brace] ($(xl) + (\x*0.25,\x*0.3)$);
\draw [] ($(xl) + (-\x*1.75,-\x*0.3)$) to [square left brace] ($(xl) + (-\x*1.75,\x*0.3)$);

\node (xi1) at ($(xl) + (-\x*1.5,\x*0)$) {$X_{1}^i$};
\draw [dashed] ($(xl) + (-\x*1.25,\x*0.25)$) -- ($(xl) + (-\x*1.25,-\x*0.25)$);
\node (xi1) at ($(xl) + (-\x*1,\x*0)$) {$X_{2}^i$};
\draw [dashed] ($(xl) + (-\x*0.75,\x*0.25)$) -- ($(xl) + (-\x*0.75,-\x*0.25)$);
\node (xidot) at ($(xl) + (-\x*0.5,\x*0)$) {$\cdots$};
\draw [dashed] ($(xl) + (-\x*0.25,\x*0.25)$) -- ($(xl) + (-\x*0.25,-\x*0.25)$);
\node (xil) at (xl) {$X_{\ell}^i$};

\draw [decorate,decoration={brace,amplitude=3pt},xshift=0pt,yshift=0pt]
($(xl) + (\x*0.25,-\x*0.6)$) -- ($(xl) + (-\x*1.75,-\x*0.6)$)  node [black,midway,yshift=-\x*0.5cm] 
{$\ell$ blocks of size $k$};


\node (eq) at ($(dd6) + (\x*0.625,\x*1.5)$) {$=$};

\coordinate (y1) at ($(dd6) + (\x*1.25,\x*1.5)$);

\draw [] ($(y1) + (\x*1.8,-\x*0.3)$) to [square right brace] ($(y1) + (\x*1.8,\x*0.3)$);
\draw [] ($(y1) + (-\x*0.25,-\x*0.3)$) to [square left brace] ($(y1) + (-\x*0.25,\x*0.3)$);

\node (xi1) at ($(y1) + (-\x*0,\x*0)$) {$r_{1}^i$};
\draw [dashed] ($(y1) + (\x*0.25,\x*0.25)$) -- ($(y1) + (\x*0.25,-\x*0.25)$);
\node (xi1) at ($(y1) + (\x*0.5,\x*0)$) {$r_{2}^i$};
\draw [dashed] ($(y1) + (\x*0.75,\x*0.25)$) -- ($(y1) + (\x*0.75,-\x*0.25)$);
\node (xidot) at ($(y1) + (\x*1,\x*0)$) {$\cdots$};
\draw [dashed] ($(y1) + (\x*1.25,\x*0.25)$) -- ($(y1) + (\x*1.25,-\x*0.25)$);
\node (yil) at ($(y1)+(\x*1.5,\x*0)$) {\hspace{5pt}$r_{\ell+1}^i$};

\draw [decorate,decoration={brace,amplitude=3pt},xshift=0pt,yshift=0pt]
($(y1) + (\x*1.75,-\x*0.6)$) -- ($(y1) + (-\x*0.25,-\x*0.6)$)  node [black,midway,yshift=-\x*0.5cm] 
{$\ell+1$ blocks of size $n$};

\end{tikzpicture}
  \caption{Illustration of the received symbols for~$M=1$.}
  \label{fig:responses}
\end{figure}

\section{Protecting against block erasures} \label{sec:block}
In the previous section, we showed how to design queries such that the symbols of the desired file recovered from the responses are symbols of a code of higher dimension and memory~$M$. While this setting asymptotically achieves the same PIR rate as a comparable system that downloads blocks without memory, it has no immediate advantages. In this section, we utilize the construction to design a PIR scheme that is able to stream files consisting of many stripes in the presence of bursts of block erasures, \emph{i.e.}, consecutive iterations where all the responses of the servers are lost. Since we are interested in streaming applications, decoding should be possible without a big delay and without querying for more data or retransmission of blocks. Therefore, we consider a sliding decoding window of~$N$ blocks and denote the maximum burst length of block erasures in a window by~$\epsilon$.

\begin{definition}\label{def:PIRBlockErasureTemp}
  Consider a $[n,k]$ storage system storing $m$ files, each divided into $\ell$ stripes of non-zero entropy. A PIR scheme is said to be $N$-window decodable in the presence of $\epsilon$-bursts of block erasures if
  \begin{enumerate}
  \item given the previous $N-1$ replies stripe $X^i_\xi$ is decodable in iteration $\xi$, i.e.,
    \begin{equation*}
      H(X_\xi^i | q, i, r_{\xi-N+1}^i,...,r_{\xi-1}^i) = 0
    \end{equation*}
  \item each burst of $\epsilon$ block erasures can be resolved within $N$ blocks, i.e.,
    \begin{equation*}
    H(X^i_{\xi-N+1},..., X^i_{\xi} |q, i, r_{\xi-N+\epsilon}^i,...,r_{\xi}^i) = 0 \ .
  \end{equation*}
\end{enumerate}
\end{definition}
The first condition ensures that under normal operation, i.e., when no block erasures occur, there is minimal delay, as the new part of the file can be decoded immediately when a new block is received. The second condition ensures that a burst of $\epsilon$ block erasures can be resolved within a window of at most $N$ blocks. To achieve this, any scheme requires at least $\epsilon$ more iterations than stripes.
\begin{lemma}
  A scheme as in Definition~\ref{def:PIRBlockErasureTemp} requires at least $\ell+\epsilon$ iterations.
\end{lemma}
\begin{IEEEproof}
  Let the number of iterations in the scheme be $\ell+\delta$. Assume the responses $r_{\ell}^i,...,r_{\ell+\min\{\delta+1,\epsilon\}-1}^i$ are erased. From Definition~\ref{def:PIRBlockErasureTemp} and the fact that conditioning
  decreases entropy it follows that
  \begin{equation*}
    H(X^i_{\ell} |q, i, r_{\ell+\epsilon}^i,...,r_{\ell+\min\{\delta,N-1\}}^i) =  0 \ ,
  \end{equation*}
  which can only be satisfied if $\delta \geq \epsilon$, since $X$ is independent of $q$ and $i$.
\end{IEEEproof}

An upper bound on the rate achievable by such a scheme is directly related to the rate achievable by a regular PIR scheme for the same storage system.
\begin{theorem}\label{thm:PIRRateBoundBlock}
  A PIR scheme as in Definition~\ref{def:PIRBlockErasureTemp} is of rate
  \begin{equation*}
    R_{\mathrm{PIR}}^b \leq \left(1-\frac{\epsilon}{N} \right) R_{\mathrm{PIR}}^{\mathrm{opt.}} \ ,
  \end{equation*}
  where $R_{\mathrm{PIR}}^{\mathrm{opt.}}$ is the optimal rate for PIR from an $[n,k]$ coded storage system with $t$-collusion.
\end{theorem}
\begin{IEEEproof}
First consider the erasure of the blocks $[l N+1 , lN+\epsilon], l \in \left[ 0, \left\lceil \frac{\ell}{N} \right\rceil-1\right]$, i.e., the erasure of the first $\epsilon$ blocks in each window of $N$ blocks, starting with the first block. By Definition~\ref{def:PIRBlockErasureTemp} all stripes can be recovered from the remaining blocks, and from the definition of the PIR capacity it follows that
\begin{equation*}
  H(X^i_1,...,X^i_{\ell}) \leq R_{\mathrm{PIR}}^{\mathrm{opt.}}  \sum_{\xi\in [\ell] \setminus [l N+1 , lN+\epsilon] } \sum_{j=1}^n H(r_{\xi,j}^i)\ .
\end{equation*}
The same holds for any $z \in [0,N-1]$ shift of the erasure pattern (see Figure~\ref{fig:erasurePatterns} for an illustration) and we denote the corresponding sets by
\begin{equation*}
\mathcal{A}_z = \bigcup_{l \in \left[ 0, \left\lceil \frac{\ell}{N} \right\rceil-1\right]} \left[lN + z + 1, lN + z+\epsilon \!\! \mod \left\lceil \frac{\ell}{N} \right\rceil N\right] \ .
\end{equation*}
Summing over all $z\in [0,N-1]$ and observing that every block/reply is erased exactly $\epsilon$ times gives
\begin{align*}
  \sum_{z=0}^{N-1} H(X^i_1,...,X^i_{\ell}) &\leq \sum_{z=0}^{N-1} R_{\mathrm{PIR}}^{\mathrm{opt.}} \sum_{\xi\in [\ell+\epsilon] \setminus \mathcal{A}_{z}} \sum_{j=1}^n H(r_{\xi,j}^i) \\
  N H(X^i_1,...,X^i_{\ell}) &\leq  R_{\mathrm{PIR}}^{\mathrm{opt.}} (N-\epsilon) \sum_{\xi=1 }^{\ell+\epsilon} \sum_{j=1}^n H(r_{\xi,j}^i) \\
  \underbrace{\frac{H(X^i_1,...,X^i_{\ell})}{\sum_{\xi=1 }^{\ell+\epsilon} \sum_{j=1}^n H(r_{\xi,j}^i)}}_{R_{\mathrm{PIR}}^b } &\leq  \left(1-\frac{\epsilon}{N} \right) R_{\mathrm{PIR}}^{\mathrm{opt.}} \ .
\end{align*}

\begin{figure}
  \centering
  \def\x{1}

\begin{tikzpicture}


\node (S1) at (0,0) [draw=none] {};


  \foreach \i in {3,4,5,8,9,10,13,14,15,18,19,20,23,24}{
    \node (r\i) at ($(S1)+(\i*\x*0.63cm,\x*2)$) [draw=none,minimum width=\x*0.7cm,minimum height=\x*0.45cm] {$r_{\i}^i$};
    }

  \foreach \i in {1,2,6,7,11,12,16,17,21,22}{
    \node (r\i) at ($(S1)+(\i*\x*0.63cm,\x*2)$) [draw=none,minimum width=\x*0.7cm,minimum height=\x*0.45cm, fill=red!30!white] {$r_{\i}^i$};
    }

    \foreach \i in {1,4,5,6,9,10,11,14,15,16,19,20,21,24}{
    \node at ($(S1)+(\i*\x*0.63cm,\x*1.2)$) [draw=none,minimum width=\x*0.7cm,minimum height=\x*0.45cm] {$r_{\i}^i$};
    }

  \foreach \i in {2,3,7,8,12,13,17,18,22,23}{
    \node at ($(S1)+(\i*\x*0.63cm,\x*1.2)$) [draw=none,minimum width=\x*0.7cm,minimum height=\x*0.45cm, fill=red!30!white] {$r_{\i}^i$};
    }

    \foreach \i in {1,2,5,6,7,10,11,12,15,16,17,20,21,22}{
    \node at ($(S1)+(\i*\x*0.63cm,\x*0.4)$) [draw=none,minimum width=\x*0.7cm,minimum height=\x*0.45cm] {$r_{\i}^i$};
    }

  \foreach \i in {3,4,8,9,13,14,18,19,23,24}{
    \node at ($(S1)+(\i*\x*0.63cm,\x*0.4)$) [draw=none,minimum width=\x*0.7cm,minimum height=\x*0.45cm, fill=red!30!white] {$r_{\i}^i$};
  }

  \foreach \i in {1,2,3,6,7,8,11,12,13,16,17,18,21,22,23}{
    \node at ($(S1)+(\i*\x*0.63cm,-\x*0.4)$) [draw=none,minimum width=\x*0.7cm,minimum height=\x*0.45cm] {$r_{\i}^i$};
    }

  \foreach \i in {4,5,9,10,14,15,19,20,24}{
    \node at ($(S1)+(\i*\x*0.63cm,-\x*0.4)$) [draw=none,minimum width=\x*0.7cm,minimum height=\x*0.45cm, fill=red!30!white] {$r_{\i}^i$};
    }

    \foreach \i in {2,3,4,7,8,9,12,13,14,17,18,19,22,23,24}{
    \node at ($(S1)+(\i*\x*0.63cm,-\x*1.2)$) [draw=none,minimum width=\x*0.7cm,minimum height=\x*0.45cm] {$r_{\i}^i$};
    }

  \foreach \i in {1,5,6,10,11,15,16,20,21}{
    \node at ($(S1)+(\i*\x*0.63cm,-\x*1.2)$) [draw=none,minimum width=\x*0.7cm,minimum height=\x*0.45cm, fill=red!30!white] {$r_{\i}^i$};
    }


    \draw [dashed, rounded corners=1pt, gray, very thick] (r1.north west) rectangle (r5.south east);

    \draw [very thick,->,gray] ($(r2)+(0,\x*0.5)$) -- ($(r4)+(0,\x*0.5)$);
    \node at ($(r3)+(0,\x*0.6)$) [draw=none,anchor=south, gray!50!black] {Decoding window};

\end{tikzpicture}
  \caption{Each row illustrates a erasure pattern correctable by Definition~\ref{def:PIRBlockErasureTemp} for $\ell=22$, $\epsilon=2$, and $N=5$. Erased blocks are indicated in red.}
  \label{fig:erasurePatterns}
\end{figure}
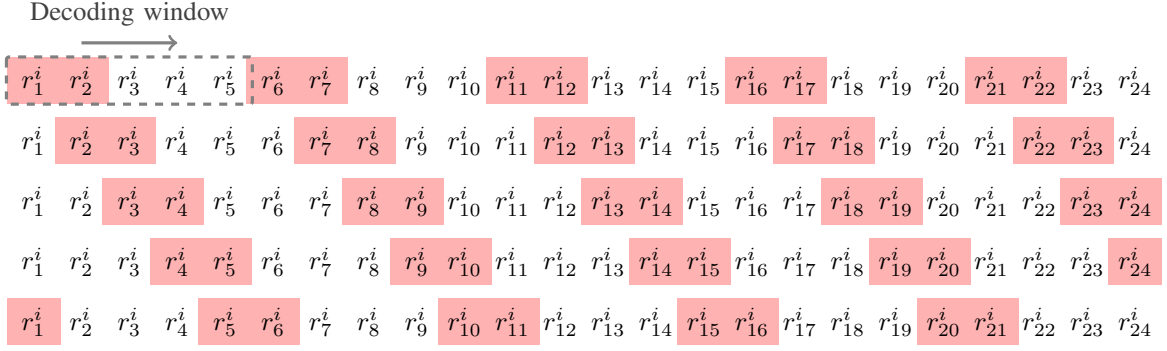
\end{IEEEproof}

\subsection{PIR scheme protecting against bursts of block erasures}
The scheme presented in the following is based on the extension of the star product scheme introduced in Section~\ref{sec:pirconv}. The protection against bursts of block erasures is achieved by increasing the number of symbols downloaded in each iteration and carefully choosing the corresponding positions in each block. To protect against these erasures, more symbols of each block have to be retrieved privately in each iteration than in the setting of the previous section.
\begin{lemma} \label{lem:khat}
  The number of symbols~$\gamma$ privately retrieved in each non-erased block has to satisfy
  \begin{equation*}
    \gamma \geq \frac{Nk}{N-\epsilon} \ .
  \end{equation*}
\end{lemma}
\begin{IEEEproof}
  Losing~$\epsilon$ consecutive blocks out of~$N$ blocks leaves~$(N-\epsilon)\frac{N k}{N-\epsilon} = Nk$ retrieved
  symbols in that window, the minimal number to recover the corresponding~$Nk$ message symbols.
\end{IEEEproof}
In the following the memory $M$ is chosen according to the number of block erasures that the system is supposed to tolerate. Trivially~$M \geq \epsilon$ has to hold, since a burst of~$M+1$ block erasures makes the received symbols independent of some stripe of the file and recovery impossible. Generally it is desirable to keep the memory low, since larger memory $M$ increases the size of the query and the computational complexity on the server and user side, so the following scheme is designed to tolerate any burst of $M$ block erasures while introducing the minimal required memory of $M=\epsilon$. Further, for ease of notation, we assume that $(d_\star-1) | \gamma$.

\subsubsection{Query}

The queries are similar to \secref{subsec:queryconv}, but by \lemref{lem:khat} it has to hold that
\begin{equation} \label{eq:cardJblock}
  |J| \geq \frac{Nk}{N-\epsilon} \ .
\end{equation}
The set~$J$ has to be chosen such that recovery of the file is possible in the presence of block erasures.

\begin{definition} \label{def:rankcond}
  Let~$G$ be the generator matrix of a convolutional code of memory~$M=\epsilon$ and~$(n,k)$ component codes generated by~$G_z$ , $z\in [0,M]$. We say that a set~$J \subset [n]$ has the \emph{recovering property} if
  \begin{equation*}
    \mathrm{rank} \left( \left.G^{\vphantom{d}} \right|^{\mathcal{R}}_{J} \right) = N k
  \end{equation*}
  for any~$\mathcal{R} = [\xi-N+\epsilon+1,\xi]; \xi \in [\ell+M]$, where~$\left.G^{\vphantom{d}} \right|^{\mathcal{R}}_{J}$ denotes the restriction of~$G$ to the positions in~$J$ in each block and to the blocks indexed by~$\mathcal{R}$.
\end{definition}
This assures that a burst of~$\epsilon$ block erasures can be recovered while still being within the window of~$N$ blocks. In our setting, the matrices $G_z$ will be generator matrices of $\code \star E_{z+1}$.
 In the Appendix, we will show that when the matrices $G_z$ generate a Reed-Solomon code, then the recovering property is equivalent to a rather simple algebraic criterion. We also show that codes with sets satisfying the recovering property exist.
\begin{remark}
In the original star product scheme, the equivalent of the set $J$ has to be an information set of the storage code. Then, when the corresponding positions of the encoded desired file are recovered, the actual file can be recovered from these positions. Here, the set $J$ should be chosen such that in the presence of block erasures, the recovered positions in the remaining blocks are an information set of the higher dimension code remaining in the current window.
\end{remark}

\subsubsection{Decoding}

Decoding the queries to obtain the respective stripes of the requested file consists of two main steps: erasure decoding to obtain the linear combination of desired symbols and recovering the stripes from these symbols.
\begin{theorem}
  Let~$J$ be a set with the recovering property as in \defref{def:rankcond}, and let~$n\geq k+t-1+|J|$. For any set~$\mathcal{R} = [\xi-N+\epsilon+1,\ldots,\xi]; \xi \in [\ell+M]$; the stripes~$\{X_{\xi-N+1}^i,\ldots,X_{\xi}^i \}$ can be recovered from the responses~$r_s^i, s\in \mathcal{R}$.
\end{theorem}
\begin{IEEEproof}
  The code~$\code \star \coded$ has distance~$\dcsd = n-(k+t-1)+1 \geq |J|+1$, and it follows that the vector
  \begin{equation*}
    \sum_{z=0}^{M} E_{z+1} \star Y_{\xi-z}^i = \sum_{z=0}^{M} X_{\xi-z}^i \cdot G_{\code \star E_{z+1}}
  \end{equation*}
  can be recovered for any~$\xi \in \mathcal{R}$. By \defref{def:rankcond}, the matrix generating these vectors has rank~$Nk$ and thus all~$N$ stripes in this window can be recovered.
\end{IEEEproof}
\subsection{Performance}
\begin{lemma}\label{lem:rateblock}
  The PIR rate is given by
  \begin{align*}
    R_{\mathrm{PIR}}^b \leq \left(1-\frac{\epsilon}{N}\right) \frac{\ell(n-(k+t-1))}{ (\ell+\epsilon)n} \ ,
  \end{align*}
  with equality for~$\gamma = \frac{Nk}{N-\epsilon}$.
\end{lemma}
\begin{IEEEproof}
  By definition,~$Nk$ information symbols have to be downloaded in each window of~$N$ blocks. In each round~$\dcsd -1\leq n-(k+t-1)$ symbols of the~$\gamma$ desired symbols in a block can be downloaded, so $\frac{\gamma}{d_\star-1}$ rounds are required (see \cite{freij2016private} for details) for each block. Hence, the total number of downloaded symbols of $\field$ is $(\ell+\epsilon)\frac{\gamma}{d_\star-1}n$ and the PIR rate is given by
  \begin{align*}
    R_{\mathrm{PIR}} &= \frac{H(X^i)}{(\ell+\epsilon)\frac{\gamma}{d_\star-1}n \log(|\field|)}\\
                    &=   \frac{\ell k}{(\ell+\epsilon)\frac{\gamma}{d_\star-1}n} \\
            &\leq \frac{\ell}{\ell+\epsilon}\frac{k(n-(k+t-1))}{\frac{Nk}{N-\epsilon}n}\\
            &=\left(1-\frac{\epsilon}{N}\right) \frac{\ell(n-(k+t-1))}{ (\ell+\epsilon)n} \ .
  \end{align*}
\end{IEEEproof}

\begin{corollary}\label{col:cap}
  Let $\gamma = \frac{Nk}{N-\epsilon}$ and $m\rightarrow \infty$. Then the PIR rate of Lemma~\ref{lem:rateblock} approaches the upper bound given in Theorem~\ref{thm:PIRRateBoundBlock} as
  \begin{equation*}
    R_{\mathrm{PIR}}^b = \left(1-\frac{\epsilon}{N}\right) \frac{\ell(n-(k+t-1))}{ (\ell+\epsilon)n} \;\; \stackrel{ \ell \rightarrow \infty}{\longrightarrow} \;\; \left(1-\frac{\epsilon}{N} \right) R_{\mathrm{PIR}}^{\mathrm{opt.}} \ ,
  \end{equation*}
  where $R_{\mathrm{PIR}}^{\mathrm{opt.}}$ is given by the (conjectured) PIR capacity for $m \rightarrow \infty$ (\cite{sun_jafar_1} for $k=t=1$, \cite{Sun2018} for $k=1, t\geq 1$, \cite{Banawan2018} for $k \geq 1, t=1$, \cite{Holzbaur2019capacity} for $k,t \geq 1$ linear PIR, and \cite{freij2016private} for conjecture on general case). 
\end{corollary}

Note that when further letting  $N \rightarrow \infty$, the PIR rate of the convolutional scheme approaches the respective (conjectured) asymptotic PIR capacities given in Corollary~\ref{col:cap}.

\begin{figure}
  \begin{subfigure}{0.5\linewidth}
    \begin{tikzpicture}
\pgfplotsset{compat = 1.3}
\begin{axis}[
	legend style={nodes={scale=0.8, transform shape}},
	width = \linewidth,
	xlabel = $N$,
	ylabel = $R_{PIR}$,
	xmin = 0,
	xmax = 30,
	ymin = 0,
	ymax = 0.25,
        ytick = {0.1,0.2},
	legend style={at={(0.54,0.5)},anchor=west}]

\addplot[color=TUMBlau,
domain = 4:30,
samples = 27,
mark=x]
{(1-3/x)*(100*(100-(75+1-1)))/((100+3)*100)};
\addlegendentry{$R_{PIR}^b$}

\addplot[color=black,
domain = 4:30,
samples = 27,
mark=triangle]
{(1-3/x)*(100-75)/100};
\addlegendentry{Upper Bound}

\end{axis}
\end{tikzpicture}

    \caption{Comparison of PIR rates for $\epsilon=3$ and different $N$.}
      \label{fig:comparisonrateblocka}
  \end{subfigure}
  \begin{subfigure}{0.5\linewidth}
    \begin{tikzpicture}
\pgfplotsset{compat = 1.3}
\begin{axis}[
	legend style={nodes={scale=0.8, transform shape}},
	width = \linewidth,
	xlabel = $N$,
	ylabel = $R_{PIR}$,
	xmin = 0,
	xmax = 30,
	ymin = 0,
	ymax = 0.25,
        ytick={0.1,0.2},
	legend pos = north east]

\addplot[color=TUMBlau,
domain = 2:30,
samples = 15,
mark=x,
mark repeat=1]
{(1-0.5)*(100*(100-(75+1-1)))/((100+x/2)*100)};
\addlegendentry{$R_{PIR}^b$}

\addplot[color=black,
domain = 2:30,
samples = 15,
mark=triangle]
{(1-0.5)*(100-75)/100};
\addlegendentry{Upper Bound}

\end{axis}
\end{tikzpicture}

    \caption{Comparison of PIR rates for $\epsilon=\frac{N}{2}$.}
      \label{fig:comparisonrateblockb}
  \end{subfigure}\\[1ex]
  \begin{center}
  \begin{subfigure}{0.5\linewidth}
    \begin{tikzpicture}
\pgfplotsset{compat = 1.3}
\begin{axis}[
	legend style={nodes={scale=0.8, transform shape}},
	width = \linewidth,
	xlabel = $\epsilon$\vphantom{N},
	ylabel = $R_{PIR}$,
	xmin = 0,
	xmax = 11,
	ymin = 0,
	ymax = 0.25,
        ytick={0.1,0.2},
	legend pos = north east]

\addplot[color=TUMBlau,
domain = 0:11,
samples = 12,
mark=x]
{(1-x/12)*(100*(100-(75+1-1)))/((100+x)*100)};
\addlegendentry{$R_{PIR}^b$}

\addplot[color=black,
domain = 0:11,
samples = 12,
mark=triangle]
{(1-x/12)*(100-75)/100};
\addlegendentry{Upper Bound}

\end{axis}
\end{tikzpicture}

    \caption{Comparison of PIR rates for $N=12$ and different $\epsilon$.}
      \label{fig:comparisonrateblockc}
  \end{subfigure}
  \end{center}
  \caption{Comparison of PIR rates for $n=100$, $k=75$, $t=1$, $\ell =100$, and $m\rightarrow \infty$. The PIR capacity for the shown parameters (coded, non-colluding) used for calculating the bound is given in~\cite{Banawan2018}. }
  \label{fig:comparisonrateblock}
\end{figure}
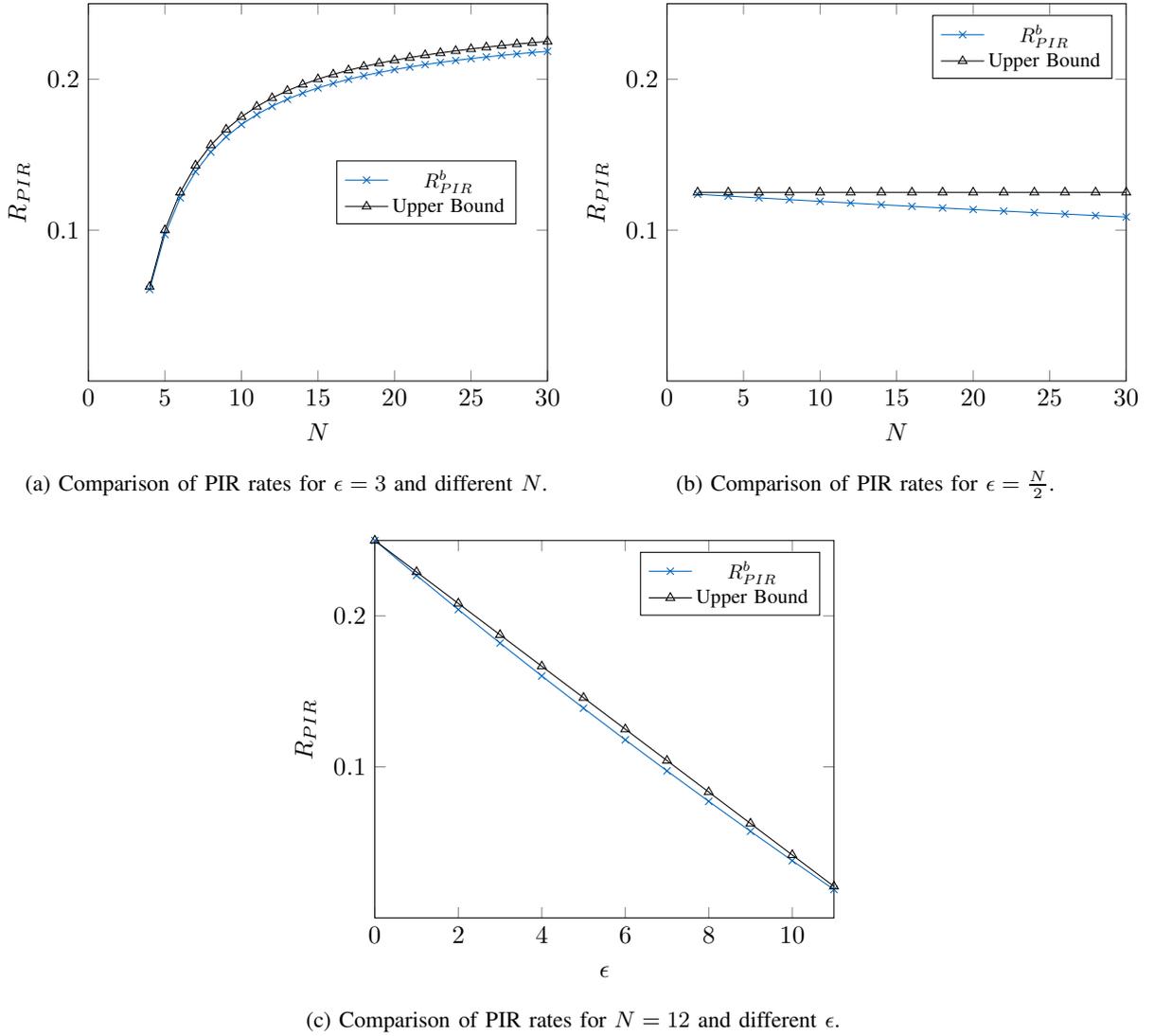

Figure~\ref{fig:comparisonrateblock} shows the upper bound on the PIR rate derived in Theorem~\ref{thm:PIRRateBoundBlock} and the rate of the proposed scheme given in Lemma~\ref{lem:rateblock} that protects against $\epsilon$ consecutive block erasures. In Figure~\ref{fig:comparisonrateblocka}, the number of consecutive erasures within a decoding window is fixed to $\epsilon=3$. 
In Figure~\ref{fig:comparisonrateblockb}, the ratio between the number of consecutive block erasures $\epsilon$ and the decoding window size~$N$ is fixed to $\epsilon = \frac{N}{2}$. For larger window size, the PIR rate of the convolutional scheme decreases gradually with the number of consecutive block erasures, since the necessary increase in memory causes an increased loss due to termination.
In Figure~\ref{fig:comparisonrateblockc}, the decoding window size is fixed to $N=12$, the loss compared to the upper bound is again solely due to termination of the code.

\begin{remark}
  Another possibility to protect against block erasures is performing the coding separate from the PIR at each node. In this case each node splits the response for each iteration into $k_s$ elements from a subfield and encodes them with an $(n_s,k_s)$ code $\code_s$ over the subfield. By the same argument as in Theorem~\ref{thm:PIRRateBoundBlock}, the rate of such a code protecting against bursts of $\epsilon$ erasures is at most $R_s \leq 1-\frac{N}{\epsilon}$, so the overall PIR rate of such a scheme is also upper bounded by $R_{PIR} \leq R_s \cdot R_{PIR}^{\mathrm{opt.}}$. This is a different approach to the problem, which leads to increased download from each node, while the approach presented in the previous section achieves a smaller download from each of the slightly larger number of nodes. However, this approach has other downsides, such as a possible increase in subpacketization, as every symbol needs to be split further to allow for the separate encoding.
\end{remark}

\subsection{Examples}

For ease of understanding, we give two examples of the described scheme for specific parameters. Example~\ref{eg:blocktrivial} shows that the window size has to be chosen sufficiently large to allow for a non-trivial scheme and a gain in PIR rate. Example~\ref{eg:blockerasure} describes each step of the scheme in detail for specific parameters and gives a class of explicit locators for which the set~$J$ has the recovering property from Definition~\ref{def:rankcond}.
\begin{example}\label{eg:blocktrivial}
  Consider the case where~$\epsilon = 1$ and~$N=2$. In this case \lemref{lem:khat} gives~$\gamma = 2k$ and the PIR rate for~$\ell \rightarrow \infty$ is~$R_{\mathrm{PIR}} = \frac{1}{2}R_{\mathrm{PIR}}^{\star}~$, where~$R_{\mathrm{PIR}}^{\star}$ is the rate achieved by the scheme in \cite{freij2016private}. In this case, the same result can be achieved with a trivial scheme that downloads each block twice.
\end{example}

\begin{example}\label{eg:blockerasure}
  Let~$m=3$,~$M=1$,~$n=6$,~$k=2$,~$t=1$,~$N=3$ and~$\epsilon=1$.  Let~$D\in \field^{6\times 6}$ be a random matrix with~$6$ i.i.d. random codewords from an~$\RS(n,t)$ code as rows and~$J = \{4,5,6\}$. Assume the user wants to retrieve the second file~$X^2$. With \eqref{eq:Evector} the query matrix is given by
  \begin{equation*}
    D +
    \left(\begin{matrix}
        0&0&0&0&0&0\\
        0&0&0&1&1&1\\
        0&0&0&0&0&0\\
        0&0&0&0&0&0\\
        0&0&0&\alpha_4^2&\alpha_5^2&\alpha_6^2\\
        0&0&0&0&0&0
      \end{matrix}\right)
    \; \in \; \field^{6\times 6} \ .
  \end{equation*}
  The query~$q_j^2$ for the~$j$-th server is given by the~$j$-th column.\\
  In the first iteration the user obtains~$X_1^2$. Now assume the second block is lost.  In the third and fourth
  iteration the nodes return~$r_{3,j}^2 = \left\langle q_j^2,(Y_{3,j},Y_{2,j})^T \right\rangle$ and
  ~$r_{4,j}^2 = \left\langle q_j^2, (Y_{4,j},Y_{3,j})^T \right\rangle$. The user receives
  \begin{align*}
    r_3^i &= \sum_{s=1}^{m} ( D_{s} \star Y_{3}^s + D_{M+s} \star Y_{2}^s) + (0,0,0,Y_{3,4}^2 + \alpha_4^2 Y_{2,4}^2, Y_{3,5}^2 + \alpha_5^2 Y_{2,5}^2, Y_{3,5}^2 + \alpha_6^2 Y_{2,6}^2) \\
    r_4 &= \sum_{s=1}^{m} ( D_{s} \star Y_{4}^2 + D_{M+s} \star Y_{3}^2) + (0,0,0,Y_{4,4}^2 + \alpha_4^2 Y_{3,4}^2, Y_{4,5}^2 + \alpha_5^2 Y_{3,5}^2, Y_{4,5}^2 + \alpha_6^2 Y_{3,6}^2) \ .
  \end{align*}
  The distance of~$\code \star \coded$ is~$\dcsd = 4$ and treating positions~$4-6$ as erasures gives
  \begin{align}
    (Y_{3,(4:6)}^2 + \alpha_{4:6}^2 \star Y_{2,(4:6)}^2, Y_{4,(4:6)}^2 + \alpha_{4:6}^2 \star Y_{3,(4:6)}^2) &=(X_2^2,X_3^2,X_4^2) \cdot
     \left(\begin{matrix}
         G_{\code \star E_2}^{4:6}& \\
         G_{\code \star E_1}^{4:6} & G_{\code \star E_2}^{4:6} \\
         &G_{\code \star E_1}^{4:6}
       \end{matrix}\right)\nonumber \\
    &= (X_2^2,X_3^2,X_4^2) \cdot
      \left(\begin{matrix}
          \alpha_4^2&\alpha_5^2&\alpha_6^2&&& \\
          \alpha_4^3&\alpha_5^3&\alpha_6^3&&& \\
          1&1&1&\alpha_4^2&\alpha_5^2&\alpha_6^2\\
          \alpha_4&\alpha_5&\alpha_6&\alpha_4^3&\alpha_5^3&\alpha_6^3\\
          &&&1&1&1\\
          &&&\alpha_4&\alpha_5&\alpha_6
        \end{matrix}\right) \ , \label{eq:matrix2}
  \end{align}
  where~$\alpha_{4:6}^2 = (\alpha_4^2,\alpha_5^2,\alpha_6^2)$.  If this matrix has full rank, the files~$X_2^2$,~$X_3^2$
  and~$X_4^2$ can be recovered. Whether it does have full rank depends on the choice of evaluation points and we will now show that we can choose evaluation points such that this matrix is invertible. Let us assume that the field size~$|\mathbb{F}| > 3$. Let~$\alpha_4, \alpha_5, \alpha_6\in\mathbb{F}$ be such that their
  squares~$\alpha_j^2$ are all distinct. Assume for a contradiction that the matrix
  \begin{align*}
    A = \left(\begin{matrix}
        \alpha_4^2&\alpha_5^2&\alpha_6^2&&& \\
        \alpha_4^3&\alpha_5^3&\alpha_6^3&&& \\
        1&1&1&\alpha_4^2&\alpha_5^2&\alpha_6^2\\
        \alpha_4&\alpha_5&\alpha_6&\alpha_4^3&\alpha_4^3&\alpha_6^3\\
        &&&1&1&1\\
        &&&\alpha_4&\alpha_5&\alpha_6
      \end{matrix}\right)
  \end{align*}
  does not have full rank, but satisfies~$xA=0$ for some non-zero row vector~$x=(x_1,\ldots , x_6)$. Denoting
  \begin{equation*}
    A'=\left(\begin{matrix}
        \alpha_4^2&\alpha_5^2&\alpha_6^2\\
        \alpha_4^3&\alpha_5^3&\alpha_6^3 \\
        1&1&1\\
        \alpha_4&\alpha_5&\alpha_6
      \end{matrix}\right)
  \end{equation*}
  and studying the first and the last three columns of~$A$ separately, we get that
  \begin{equation*}
    (x_1,\ldots, x_4) A' = (x_3,\ldots, x_6) A' = 0 .
  \end{equation*}
  As~$A'$ is a Vandermonde matrix, any three of its rows are independent, so~$x'A'=0$ implies that~$x'$ is either the
  zero vector or has full support. As we know that~$x=(x_1,\ldots , x_6)$ is not the zero vector, it follows that~$x_1$
  is also non-zero, and after scaling we may assume that~$x_1=1$. As~$A'$ has a one-dimensional left null space that
  contains both $(x_1,\ldots , x_4)$ and~$(x_3,\ldots , x_6)$, we must have~$(x_3,\ldots ,x_6)=t(x_1,\ldots ,x_4)$ for
  some $t\in \mathbb{F}$. We can therefore write \[ (x_1,x_2,x_3,x_4)=(1,s,t,ts) \] for some~$s,t\in \mathbb{F}_q$. The linear
  system of
  equations
  \begin{equation*} (1,s,t,ts)\left(\begin{matrix}
        \alpha_4^2&\alpha_5^2&\alpha_6^2\\
        \alpha_4^3&\alpha_5^3&\alpha_6^3 \\
        1&1&1\\
        \alpha_4&\alpha_5&\alpha_6
      \end{matrix}\right) = 0
  \end{equation*}
  implies that
  \begin{equation*}
    0 = \alpha_j^2 + s \alpha_j^3 + t + ts\alpha_j = (\alpha_j^2 +t)(1+ s\alpha_j)
  \end{equation*}
  holds for
  $j=4,5,6$. But since~$\alpha_j^2$ were distinct for different~$j$, at most one of the points may satisfy $\alpha_j^2 + t=0$, and at most one of them may satisfy~$1+ s\alpha_j=0$. This is a contradiction and it follows that any set of locators with distinct squares has the recovering property.
  By \eqref{eq:pirrate} the PIR rate for ~$\ell \rightarrow \infty$ is given by
  \begin{equation*}
    R_{\mathrm{PIR}}^b = \frac{2}{3} \cdot \frac{6-2}{6} = \frac{4}{9} \ .
  \end{equation*}
  For these parameters, the PIR rate of the trivial scheme is given by
  \begin{equation*}
      R_{\mathrm{PIR}} = \frac{1}{2} \cdot \frac{6-2}{6} = \frac{2}{6} < R_{\mathrm{PIR}}^b \ .
  \end{equation*}
\end{example}

\section{PIR with Byzantine Servers and convolutional codes}~\label{sec:byzantine}
In this section, we consider incorrectly received responses, due to either Byzantine servers or errors during transmission. We focus on constructions that result in a convolutional code of memory~$M=1$, \emph{i.e.}, UM codes.
For these codes, the decoder introduced in \cite{Dettmar1995} can efficiently decode up to half the designed extended row distance, by a combination of \emph{bounded minimum distance} (BMD) decoding in the blocks and trellis-based decoding with the Viterbi algorithm. A key step in this algorithm is decoding blocks in the cosets given by successfully decoded neighboring blocks. It is therefore imperative for a good performance to design the code such that these cosets have good distance properties. In the following, we describe a scheme that achieves this goal in the PIR setting.

\subsection{Query}

We query for two stripes in each block (\emph{i.e.}, unit memory~$M=1$) and design the queries such that when one block can be decoded and both stripes can be recovered, the neighboring blocks have good distance properties in the corresponding cosets.

Let~$D \in \field^{2m \times n}$ be as in \eqref{eq:query} and~$\coded$ be an~$\RS(n,t)$ code. The query for the~$j$-th server is given by
\begin{equation}
  \label{eq:querybyz}
  q_j^i = D_{\cdot,j} + e_i E_1 + e_{m+i} E_2 \ ,
\end{equation}
where~$E_1 = (a_j^{-k})$,~$E_2 = (a_j^{k+t-1})$ and~$e_i$ is the~$i$-th standard basis vector.

\subsection{Response}

The response to one query consists of~$\ell + 1$ parts. In iteration~$\xi$ the response of server~$j$ is given by
\begin{equation}
  r_{\xi,j}^i = \left\langle q_j^i , (Y_{\xi,j},Y_{\xi-1,j})^T \right\rangle \ ,
\end{equation}
where~$Y_0 = Y_{\ell+1} = 0$ and~$Y_{\xi}=X_{\xi}G$ denotes the matrix storing the~$\xi$-th part of every file.

\subsection{Decoding}

The user receives
\begin{equation*}
  r_{\xi}^i = \underbrace{\sum_{s=1}^m (D_{s} \star Y_{\xi}^s  + D_{m+s} \star Y_{\xi-1}^s)}_{\in \code \star \coded} + \underbrace{E_1 \star Y_{\xi}^i}_{\in \code \star E_1} + \underbrace{E_2 \star Y_{\xi-1}^i}_{\in \code \star E_2} + w_\xi  \ ,
\end{equation*}
where~$w_\xi$ denotes the error vector of iteration~$\xi$.

\begin{lemma}
  \label{lem:distancesbyz}
  The codes~$\code \star (\coded + E_1 + E_2)$,~$\code \star (\coded + E_1)$, and~$\code \star (\coded + E_2)$ have
  respective distances~$d_{\code \star (\coded+E_1+E_2)} = n-3k-t+2$ and
 ~$d_{\code \star (\coded +E_1)} = d_{\code \star (\coded +E_2)} = n-2k-t+2$. The codes~$\code \star \coded$,
 ~$\code \star E_1$ and~$\code \star E_2$ intersect trivially.
\end{lemma}
\begin{IEEEproof}
  An~$\RS(n,k,1)$ code is the evaluation of all polynomials~$f(z)$ with~$\deg(f(z)) \leq k-1$ at the evaluation
  points~$\alpha_j$. Multiplying any polynomials corresponding to the codes~$\code, \coded, E_1$ and~$E_2$ gives
  \begin{align*}
    f_{\code}(z) \cdot (f_\coded(z) + u_{-k}' z^{-k} + u_{k+t-1}' z^{k+t-1})
    &= \underbrace{\sum_{\iota=0\vphantom{k}}^{k+t-2} u_\iota z^\iota}_{\in\code \star \coded} +  \underbrace{\sum_{\iota=-k}^{-1} u_\iota z^\iota}_{\in\code \star E_1} + \underbrace{\sum_{\iota=k+t-1}^{2k+t-2} u_\iota z^\iota}_{\in\code \star E_2}\\
    &= z^{-k}  \sum_{\iota=0}^{3k+t-2} u_{\iota-k} z^\iota \ ,
  \end{align*}
  where~$u_\iota \in \field$. Evaluating this polynomial at~$\alpha_j ,j\in [n]$, gives a codeword of~$\code \star (\coded + E_1 + E_2) = \RS(n,3k+t-1,(\alpha_j^{-k}))$. By the same argument, it holds that~$\code \star (\coded +E_1) = \RS(n,2k+t-1,(\alpha_j^{-k}))$ and~$\code \star (\coded +E_2) = \RS(n,2k+t-1,1)$. The distances follow from the Singleton bound and the trivial intersection from the distinct powers in the polynomials.
\end{IEEEproof}

To illustrate we give an example for explicit parameters.
\begin{example} \label{eg:byzmatrix}
  Let $n= 10$, $k=2$ and $t=2$. The defined matrices are given by:
  \begin{align*}
    G_\code &= G_\coded = \left(
    \begin{array}{ccccc}
      1&1&\cdots&1&1 \\
      \alpha_1&\alpha_2&\cdots&\alpha_9&\alpha_{10}
    \end{array} \right) ,\\
    E_1 &= \left(
    \begin{array}{ccccc}
      \alpha_1^{-2}&\alpha_2^{-2}&\cdots&\alpha_9^{-2}&\alpha_{10}^{-2}
    \end{array} \right) , \quad \quad
    E_2 = \left(
    \begin{array}{ccccc}
      \alpha_1^{3}&\alpha_2^{3}&\cdots&\alpha_{9}^{3}&\alpha_{10}^{3}
    \end{array} \right) \ .
  \end{align*}
  \begin{center}
    \begin{tikzpicture}

    \node (G) {$G_{\code \star (\coded + E_1+ E_2)} = $};

  \matrix (m)[
    matrix of math nodes,
    nodes in empty cells,
    left delimiter = {(},
    right delimiter = {)},
    minimum width = 1cm 
    ] at ($(G.east)+ (0.4,0)$)  [anchor = west] {
    \alpha_1^{-2}    & \alpha_2^{-2}  & \cdots & \alpha_{9}^{-2} & \alpha_{10}^{-2} \\
    \alpha_1^{-1}    & \alpha_2^{-1}  & \cdots & \alpha_{9}^{-1} & \alpha_{10}^{-1} \\
    1^{\vphantom{1}}    & 1 & \cdots & 1 & 1^{\vphantom{1}} \\
    \alpha_1^{\vphantom{1}}    & \alpha_2^{}  & \cdots & \alpha_{9}^{} & \alpha_{10}^{\vphantom{1}} \\
    \alpha_1^{2}    & \alpha_2^{2}  & \cdots & \alpha_{9}^{2} & \alpha_{10}^{2} \\
    \alpha_1^{3}    & \alpha_2^{3}  & \cdots & \alpha_{9}^{3} & \alpha_{10}^{3} \\
    \alpha_1^{4}    & \alpha_2^{4}  & \cdots & \alpha_{9}^{4} & \alpha_{10}^{4} \\
  } ;

    \draw[thick,densely dotted,green] (m-1-1.north west) rectangle (m-5-5.south east);
    \draw[thick,dashed,blue] (m-3-1.north west) rectangle (m-7-5.south east);

    \draw[thick,densely dotted,green] ($(m-1-5.south east) + (1.5,0.05)$) -- ($(m-1-5.south east) + (2,0.05)$);
    \node at ($(m-1-5.south east) + (2.1,0)$) [anchor=west] {$G_{\code \star ( \coded +E_1)}$};
    \draw[thick,dashed,blue] ($(m-2-5.south east) + (1.5,0.05)$) -- ($(m-2-5.south east) + (2.0,0.05)$);
    \node at ($(m-2-5.south east) + (2.1,0)$) [anchor=west] {$G_{\code \star ( \coded +E_2)}$};

    \draw[fill=orange!20,draw=none] ($(m-3-5.south east) + (1.5,0.15)$) rectangle ($(m-3-5.south east) + (2,-0.05)$);
    \node at ($(m-3-5.south east) + (2.1,0)$) [anchor=west] {$G_{\code \star E_1}$};
    \draw[fill=red!20,draw=none] ($(m-4-5.south east) + (1.5,0.15)$) rectangle ($(m-4-5.south east) + (2,-0.05)$);
    \node at ($(m-4-5.south east) + (2.1,0)$) [anchor=west] {$G_{\code \star \coded }$};
    \draw[fill=yellow!20,draw=none] ($(m-5-5.south east) + (1.5,0.15)$) rectangle ($(m-5-5.south east) + (2,-0.05)$);
    \node at ($(m-5-5.south east) + (2.1,0)$) [anchor=west] {$G_{\code \star E_2}$};
    \begin{scope}[on background layer]
      \draw[fill=orange!20,draw=none] (m-1-1.north west) rectangle (m-2-5.south east);
      \draw[fill=red!20,draw=none] (m-3-1.north west) rectangle (m-6-5.south east);
      \draw[fill=yellow!20,draw=none] (m-6-1.north west) rectangle (m-7-5.south east);
    \end{scope}

\end{tikzpicture}
\end{center}
The matrix $G_{\code \star (\coded + E_1 + E_2)}$ is a generator matrix of an $\RS(10,7,(\alpha_i^{-2}))$ code. The
matrices $G_{\code \star (\coded + E_1)}$ and $G_{\code \star (\coded + E_2)}$ are generator matrices of an
$\RS(10,5,(\alpha_i^{-2}))$ and $\RS(10,5,1)$ code respectively. Further, by the linear independence of the rows of
Vandermonde matrices, it can be seen that the codes $\code \star E_1$, $\code \star E_2$ and $\code \star \coded$ intersect trivially. %
\end{example}

\begin{remark}
The general approach presented here is not necessarily limited to GRS codes, however, as illustrated in Example~\ref{eg:byzmatrix}, we require that the sum of multiple codes, only differing in their column multipliers, is again a code of large distance. This very specific property further motivates our limitation to GRS codes in this work.
\end{remark}

The large number of states makes trellis decoding of the convolutional code infeasible. In \cite{Dettmar1995} an algorithm combining BMD decoding in the blocks and Viterbi decoding on a reduced trellis is given, with decoding complexity only cubic in~$n$, if the complexity of the block decoders is quadratic in~$n$. We give a brief description of this algorithm and show how it can be applied to decode the responses.
\begin{enumerate}
\item Decode each received block in~$\code_\alpha = \code \star ( \coded + E_1 + E_2)$, an~$\RS(n,3k+t-1)$ code of distance~$d_\alpha = n-3k-t+2$.
\item From the blocks successfully decoded in step 1) decode~$l_F$ steps forward and~$l_B$ backward (see \cite{Dettmar1995}) in the respective coset~$\code \star (\coded + E_1)$ or~$\code \star (\coded + E_2)$. By \lemref{lem:distancesbyz} these are~$\RS(n,2k+t-1)$ codes and can therefore be decoded up to half their minimum distance~$d_1 = d_2 = n-2k-t+2$.
\item Build a reduced trellis and find the maximum-likelihood path with the Viterbi algorithm.
\item By \lemref{lem:distancesbyz}, the codes~$\code \star \coded$,~$\code \star E_1$ and~$\code \star E_2$ intersect trivially, and it follows that the parts of the file~$X^i$ can be recovered uniquely from the codeword corresponding to the most likely path.
\end{enumerate}

\begin{theorem}\label{thm:decodingCapability}
  If \eqref{eq:decodingsuccess} holds, where~$\bar{d}_\iota^r$ is given by \eqref{eq:designedextrowdist} with~$d_\alpha = n-3k-t+2$ and~$d_1 = d_2 = n-2k+t+2$, decoding of the responses is successful and the file~$X^i$ is decoded correctly.
\end{theorem}
\begin{IEEEproof}
  By \cite{Dettmar1995} the maximum likelihood path will be in the reduced trellis if \eqref{eq:decodingsuccess} holds, which depends on the distance~$d_\alpha$ in each block and the distances~$d_1$ and~$d_2$ in the corresponding cosets of the neighboring blocks. For the code given by the responses~$\{r_1^i,\ldots, r_{\ell+1}^i \}$ these are shown in \lemref{lem:distancesbyz}. If the path is contained in the trellis, the Viterbi decoder will find it, as it is an ML decoder.
\end{IEEEproof}

This results guarantees that error patterns which fulfill the given conditions on the error distribution are decodable. 

\begin{corollary}
  The PIR rate of the scheme is
  \begin{equation*}
    R_{\mathrm{PIR}} = \frac{\ell k}{(\ell+1) n} \ ,
  \end{equation*}
  with~$n>3k+t-1$ and it has error correction capability similar to an optimal~$(n-(k+t-1),k)$ UM-code.
\end{corollary}
Note that the decoder introduced in \cite{Dettmar1995} and thereby the decoder presented here can also decode error patterns which are not covered by the given guarantee, as discussed in \cite{Dettmar1995}. As the evaluation of the true decoding performance of this UM-code decoder beyond the given guarantees, \emph{i.e.}, without the zero-error probability requirement, relies on simulations, it cannot be directly related to the achievable PIR rate by an analytic expression. However, the result can be related to the scheme of \cite{tajeddine2018byzantine}, where the error correction is similar to an (MDS) block code of shorter length. Similarly, the codes considered in our decoder also have the same error correction capability as a shorter (MDS) block code. Hence, the error correction capability of the presented scheme compares to that of the scheme in \cite{tajeddine2018byzantine} the same as that of an MDS block code to the UM-code decoding in \cite{Dettmar1995}, \emph{i.e.}, in any non-private setting where a block convolutional code performs better than a comparable block code, our scheme will perform better when the privacy requirement is introduced.

\subsection{Combination of Block Erasures and Byzantine servers}

A combination of the presented schemes which protects against both considered error models, \emph{i.e.}, block erasures and Byzantine servers/channel errors, is not directly possible as the methods used to recover the symbols of the desired file are different (erasure decoding vs. trivially intersecting codes). However, it is possible to give conditions under which the scheme for Byzantine servers also protects against single block erasures, \emph{i.e.}, the case of~$M=\epsilon=1$. By observing that the algorithm of~\cite{Dettmar1995} and therefore also our algorithm do not require termination, it is apparent that if the condition for successful decoding given in~\eqref{eq:decodingsuccess} is fulfilled between any two erasures, the erased blocks can also be recovered from the correctly decoded neighboring blocks. For a higher number of consecutive block erasures this is not possible, as the decoding algorithm of~\cite{Dettmar1995} is designed only for unit memory codes, \emph{i.e.}, memory $M=1$, and trivially $\epsilon\leq M$ has to hold. The generalization of the decoding algorithm for higher memory, and therefore increased protection against block erasures, and its application to the PIR setting are open problems.

\section{Conclusion}

In this paper, we have considered information-theoretical private streaming by combining the star product PIR scheme~\cite{freij2016private} with a block convolutional structure, thereby introducing the known benefits of codes with memory into the decoding of privately streamed/downloaded data. We introduced two schemes for different channels, \emph{i.e.}, a block erasure channel and a non-bursty channel (\emph{e.g.,} AWGN), that are suitable for streaming/downloading files, when the file size is larger than the packet size communicated in each iteration. Both work on the same database and the user can adapt to changing channel conditions by designing queries accordingly. Further, the PIR rates of both schemes are derived and compared to those of known schemes.

Future work includes the combination of the two schemes and design of an additional outer code to improve the error-correction performance.

\section*{Acknowledgment}
The authors would like to acknowledge Oliver W. Gnilke and Sven Puchinger for fruitful discussions on this topic
and for helpful comments regarding the manuscript.

\bibliographystyle{IEEEtran}
\bibliography{IEEEabrv,pir,references_cami,coding1,coding2}

%

\section*{Appendix}

\section*{Proof of existence}
In this section, we explore the existence of locators with the recovering property of Definition~\ref{def:rankcond}, when the block matrices are generator matrices of a Reed-Solomon code. First we formally define the matrix obtained by restricting the generator matrix of a convolutional code to the
$N-\epsilon$ non-erased blocks and $|J|=\gamma$ code positions. We set $\epsilon=M$, as this is the largest number of erased blocks that can be corrected by the convolutional code and consider decoding windows of length $N=2M+1$. For simplicity, we permute rows and columns to give the matrices $G_i$ in ascending order in each block. Note that this does not change the rank of the matrix.

\begin{definition}\label{def:Amat}

  Let~$V_i$ be the diagonal matrix with the entries~$(\alpha_j^{ik})_{1\leq j\leq \gamma}$ and
  \begin{align*}
    G_i = \left(\begin{matrix}
        1&1^{\vphantom{k-1}}&\cdots&1 \\
        \alpha_1&\alpha_2^{\vphantom{k-1}}&\cdots&\alpha_{\gamma} \\
        \vdots&\vdots&\ddots&\vdots \\
        \alpha_1^{k-1}&\alpha_2^{k-1}&\cdots&\alpha_{\gamma}^{k-1} \\
      \end{matrix} \right) \cdot V_{i-1} \quad \in \field^{k\times \gamma}.
  \end{align*}
  Define
  \begin{equation*}
    A = \left(
      \begin{array}{cccc}
        G_1&&& \\
        G_2&G_1&&\\
        \vdots&\vdots&\ddots& \\
        G_{M+1}&G_{M}&\hdots\vphantom{\vdots}&G_1\\
           &G_{M+1}& \hdots\vphantom{\vdots}& G_2\\
           &&\ddots&\vdots\\
           &&&G_{M+1}
      \end{array} \right) .
  \end{equation*}
\end{definition}

Now, $\{\alpha_j\}_{j=1,\dots,\gamma}$ has the recovering property if and only if $A$ has rank $Nk=(2M+1)k$. The rank of $A$ can be related to the dimension of intersections of different generalized Reed-Solomon codes.
\begin{lemma} \label{lem:intersectiongeneral}
  For $A$ as in Definition~\ref{def:Amat} it holds that
  \begin{equation*}
    \rk (A) = (2M+1)k \quad \Longleftrightarrow \quad \left\langle G_1 \right\rangle + \sum_{i=1}^{M} (\left\langle G_{1} \right\rangle \cap \left\langle G_{-M} \right\rangle )V_i \;\;\text{is a direct sum.}
  \end{equation*}
\end{lemma}
\begin{IEEEproof}
  Multiplying~$A$ from the right by a transformation matrix gives
  \begin{equation*}
    \label{eq:Atrans}
    A'  = \left(
      \begin{smallmatrix}
        G_1&&& \\
        G_2&G_1&&\\
        \vdots&\vdots&\ddots& \\
        G_{M+1}&G_{M}&\hdots\vphantom{\vdots}&G_1\\
           &G_{M+1}& \hdots\vphantom{\vdots}& G_2\\
           &&\ddots&\vdots\\
           &&&G_{M+1}
      \end{smallmatrix} \right) \cdot \left(
    \begin{smallmatrix}
      -V_{-M} &&&& \\
      V_{-M+1} & -V_{-M+1}& && \\
      &V_{-M+2} & \ddots && \\
      &&\ddots &-V_{-1}&\\
      &&&V_0&V_0
    \end{smallmatrix}\right)  = \left(
      \begin{smallmatrix}
        -G_{-M+1}&0&\cdots&0&0\\
        0&-G_{-M+2}&\hphantom{G_{M+1}}&0&0\\
        \vdots & & \ddots &  &\vdots \\
        0&0&\cdots& -G_0 & 0\\
        0&0&&0&G_1\\
        G_2 & 0 &  & 0 & G_2\\
        0 & G_3 &\cdots & 0 & G_3\vphantom{\ddots}\\
        \vdots & & \ddots & & \vdots\\
        0 & 0 &  & 0&G_M \\
        0 & 0 & \cdots & G_{M+1} & G_{M+1} \vphantom{\ddots}
      \end{smallmatrix} \right).  \end{equation*}
  The full rank of the transformation matrix follows from its upper diagonal structure, so $\rk(A)=\rk(A')$. Now
  $\rk(A') < (2M+1)k$ if and only if there exists an $x$ such that $x\cdot A' = 0$.

  The last~$\gamma$ columns of the equation $xA'=0$ now reads
\begin{equation}\label{eq:condgeneral}
  0 = y_1 G_1 + y_2 G_2 + \cdots + y_{M+1} G_{M+1},
\end{equation}
  where~$y_i = x_{(M+i-1)k+1:(M+i)k}$.
  The first $M\gamma$ columns show that $y_i G_i\in \langle G_{i-M-1}\rangle$ for~$i\geq 2$, so writing $c_i=y_i G_i$ we get a nontrivial solution to the equation $0=c_1 + c_2 + \cdots + c_{M+1}$, where $c_1\in\langle G_1\rangle$ and $c_i\in \langle G_{i}\rangle \cap \langle G_{i-M-1}\rangle= (\langle G_{1}\rangle \cap \langle G_{-M}\rangle)V^i$ for $i\geq 2$.
  It follows that if
 ~$\left\langle G_1 \right\rangle + \sum_{i=1}^{M} (\left\langle G_{1} \right\rangle \cap \left\langle G_{-M}
  \right\rangle )V_i~$ is not a direct sum, there is a linear combination of vectors from the respective subspaces such
  that~\eqref{eq:condgeneral} is fulfilled and therefore a vector~$x$ with~$x\cdot A' = 0$ exists. If it is a direct sum, the only solution of~\eqref{eq:condgeneral} is~$y_1 = y_2 = \cdots = y_{M+1} = 0$ and therefore~$x=0$.
\end{IEEEproof}

We now give an explicit method to choose locators with the recovering property for arbitrary~$M=\epsilon$ and~$k$, where~$N=2M+1$ and $\frac{2M+1}{M+1}k\leq\gamma\leq N-k$. In particular, we show that a field with such locators always exists.

\begin{lemma}\label{lm:submatrix}
  Let~$\{\alpha_j\}_{j\in[\gamma]}\subseteq\field$ be a set of locators with~$\ord(\alpha_j) | Mk+\gamma$ for all $j\in [\gamma]$. Then for~$A$ as in Definition~\ref{def:Amat} it holds that $\rk(A)=Nk$ if and only if the set $$\{(\alpha_1^{ik-j},\dots , \alpha_\gamma^{ik-j}): 2\leq i\leq M+1, 1\leq j\leq k-\gamma\}\cup \{(\alpha_1^{j},\dots , \alpha_\gamma^{j}): 0\leq j<k\}$$ is linearly independent.
\end{lemma}
\begin{IEEEproof}
As the locators were chosen such that $\alpha_j^{Mk+\gamma}=1$, we get that $G_i + G_{i-M-1}$ is generated by the vectors \begin{align*}
    &\{(\alpha_1^{(i-1)k+j},\dots,\alpha_\gamma^{(i-1)k+j}): 0\leq j<k\}\cup\{(\alpha_1^{(i-M-2)k+j},\dots , \alpha_\gamma^{(i-M-2)k+j}): 0\leq j<k\}\\
    =&\{(\alpha_1^{(i-1)k+j},\dots ,\alpha_\gamma^{(i-1)k+j}): 0\leq j<k\}\cup\{(\alpha_1^{(i-2)k+j+\gamma}, \dots , \alpha_\gamma^{(i-2)k+j+\gamma}): 0\leq j<k\},
\end{align*} which is the set of evaluation vectors of $\gamma$ consecutive powers, and thus an independent set of vectors.
It follows that $\langle G_i\cap G_{i-M-1}\rangle$ is generated by the intersection
\begin{align}
 & \{(\alpha_1^{(i-1)k+j},\dots , \alpha_1^{(i-1)k+j}): 0\leq j<k\}\cap\{(\alpha_1^{(i-2)k+j+\gamma},\dots , \alpha_\gamma^{(i-2)k+j+\gamma}): 0\leq j<k\}\nonumber\\
 =&\{(\alpha_1^{ik-j},\dots , \alpha_\gamma^{ik-j}): 1\leq j\leq 2k-\gamma\}.
\end{align}
Observe that $$\{(\alpha_1^{j},\dots , \alpha_\gamma^{j}): 0\leq j<k\}$$ are the rows of $G_1$. Thus, by Lemma~\ref{lem:intersectiongeneral}, we have $\rk(A)=Nk$ if and only if the sum $$\langle\{(\alpha_1^{j},\dots , \alpha_\gamma^{j}): 0\leq j<k\}\rangle + \sum_{i=2}^{M+1} \langle\{(\alpha_1^{ik-j},\dots , \alpha_\gamma^{ik-j}): 1\leq j\leq 2k-\gamma\}\rangle$$ is direct, or in other words if the union of their generating sets is linearly independent.
\end{IEEEproof}

We are now ready to describe an explicit construction of regenerating sets, in the sense of Definition~\ref{def:Amat}, over suitable fields.
\begin{theorem}\label{thm:regset}
  Let $\sigma=\alpha_1\in\field$ be an element of order $Mk+\gamma$, and let $\alpha_i=\sigma^i$ for $1\leq i\leq \gamma$. Then $\rk(A)=Nk$, where $A$ is as defined in Definition~\ref{def:Amat}.
\end{theorem}
\begin{IEEEproof}
  By Lemma~\ref{lm:submatrix}, we need to show that the vectors \begin{align*}
    &\{(\alpha_1^{ik-j},\dots , \alpha_\gamma^{ik-j}): 2\leq i\leq M+1, 1\leq j\leq 2k-\gamma\}\cup \{(\alpha_1^{j},\dots , \alpha_\gamma^{j}): 0\leq j<k\}\\
    =&\{(\sigma^{ik-j},\dots , \sigma^{\gamma (ik-j)}): 2\leq i\leq M+1, 1\leq j\leq 2k-\gamma\}\cup \{(\sigma^{j},\dots , \sigma^{\gamma j}): 0\leq j<k\}
  \end{align*} are linearly independent. This set of vectors can also be written as $$\left\{(x,x^2,\dots ,x^\gamma) : x\in\left\{\sigma^{ik-j}:2\leq i\leq M+1, 1\leq j\leq 2k-\gamma\right\}\cup\left\{\sigma^{j} :0\leq j<k \right\}\right\},$$ and since $\sigma$ was chosen so that $\sigma^j$ takes different values for all $0\leq j<Mk+\gamma$, these are indeed $M(2k-\gamma)+k$ different vectors of the form $(x,x^2,\dots, x^\gamma)$. We know that any set of $\leq\gamma$ such vectors are independent, and since $$M(2k-\gamma)+k=k(2M+1)-\gamma M\leq k(2M+1)\left(1-\frac{M}{M+1}\right)=\frac{k(2M+1)}{M+1}\leq\gamma,$$ the vectors are indeed linearly independent.
\end{IEEEproof}

To apply Theorem~\ref{thm:regset} directly for an explicit construction, we need to work over a field $\field_q$ with elements of multiplicative order $Mk+\gamma$, so $Mk + \gamma | q-1$.
In the case of unit memory, we have larger flexibility to choose the field size, in that we only need $\gamma | q-1$. Therefore, we will study the unit memory case next. However, we will see in Table 1 that random assignments are very likely to yield regenerating sets even over fields without these assumptions.

\begin{theorem}\label{thm:fullrankA}
  Let~$M=1$,~$\gamma = \frac{3}{2} k$ 
  be an integer and
  ~$\ord(\alpha_i) | \gamma \; \forall \, i \in [\gamma]$. Then for~$A$ as in Definition~\ref{def:Amat} it
  holds that
  \begin{equation*}
    \rk (A) = 3k.
  \end{equation*}
\end{theorem}
\begin{IEEEproof}
  By Lemma~\ref{lem:intersectiongeneral} it holds that~$\rk (A) = 3k$ if
 ~$\left\langle G_1 \right\rangle + (\left\langle G_2 \right\rangle \cap \left\langle G_{0} \right\rangle)$ is a direct
  sum, which is equivalent to
  \begin{equation*}
    \dim(\left\langle G_0 \right\rangle \cap \left\langle G_1 \right\rangle \cap \left\langle G_2 \right\rangle) = \dim(\left\langle G_1 \right\rangle \cap \left\langle G_2 \right\rangle \cap \left\langle G_{3} \right\rangle) = 0 .
  \end{equation*}
  The matrices~$G_1$ and~$G_2$ are given by
  \begin{align*}
    G_1 &= \left(\begin{matrix}
        1&1^{\vphantom{k-1}}&\cdots&1 \\
        \alpha_1&\alpha_2^{\vphantom{k-1}}&\cdots&\alpha_{\gamma} \\
        \vdots&\vdots&\ddots&\vdots \\
        \alpha_1^{k-1}&\alpha_2^{k-1}&\cdots&\alpha_{\gamma}^{k-1} \\
      \end{matrix} \right) \\
    G_2 &= \left(\begin{matrix}
        \alpha_1^k&\alpha_2^{k}&\cdots&\alpha_{\gamma}^k \\
        \vdots&\vdots&\ddots&\vdots \\
        \alpha_1^{\gamma-1}&\alpha_2^{\gamma-1}&\cdots&\alpha_{\gamma}^{\gamma-1} \\
        \alpha_1^{\gamma}&\alpha_2^{\gamma}&\cdots&\alpha_{\gamma}^{\gamma} \\
        \vdots&\vdots&\ddots&\vdots \\
        \alpha_1^{2k-1}&\alpha_2^{2k-1}&\cdots&\alpha_{\gamma}^{2k-1} \\
      \end{matrix} \right) \stackrel{(a)}{=} \left(\begin{matrix}
        \alpha_1^k&\alpha_2^{k}&\cdots&\alpha_{\gamma}^k \\
        \vdots&\vdots&\ddots&\vdots \\
        \alpha_1^{\gamma-1}&\alpha_2^{\gamma-1}&\cdots&\alpha_{\gamma}^{\gamma-1} \\
        1&1&\cdots&1 \\
        \vdots&\vdots&\ddots&\vdots \\
        \alpha_1^{\frac{1}{2}k-1}&\alpha_2^{\frac{1}{2}k-1}&\cdots&\alpha_{\gamma}^{\frac{1}{2}k-1}
      \end{matrix} \right)
  \end{align*}
  where~$(a)$ holds because~$\ord(\alpha_i) | \gamma$. As
  ~$\left\langle G_1 \right\rangle \cup \left\langle G_2 \right\rangle$ contains all rows of a~$\gamma \times \gamma$
  Vandermonde matrix it spans the entire space~$\field^\gamma$ and therefore
  ~$\dim (\left\langle G_1 \right\rangle \cap \left\langle G_2 \right\rangle) \leq 2k-\gamma$. It follows that a
  complete basis of the intersection is given by
  \begin{equation*}
    \mathrm{basis}(\left\langle G_1 \right\rangle \cap \left\langle G_2 \right\rangle) = \left(
      \begin{matrix}
        1&1&\cdots&1 \\
        \vdots&\vdots&\ddots&\vdots \\
        \alpha_1^{\frac{1}{2}k-1}&\alpha_2^{\frac{1}{2}k-1}&\cdots&\alpha_{\gamma}^{\frac{1}{2}k-1}
      \end{matrix} \right).
  \end{equation*}
  By the same argument
  \begin{equation*}
    \mathrm{basis}(\left\langle G_2 \right\rangle \cap \left\langle G_3 \right\rangle) = \left(
      \begin{matrix}
        \alpha_1^{\frac{1}{2}k}&\alpha_2^{\frac{1}{2}k}&\cdots&\alpha_{\gamma}^{\frac{1}{2}k} \\
        \vdots&\vdots&\ddots&\vdots \\
        \alpha_1^{k-1}&\alpha_2^{k-1}&\cdots&\alpha_{\gamma}^{k-1}
      \end{matrix} \right)
  \end{equation*}
  and by the linear independence of the rows of Vandermonde matrices it follows that
  \begin{equation*}
    \dim((\left\langle G_1 \right\rangle \cap \left\langle G_2 \right\rangle ) \cap (\left\langle G_2 \right\rangle \cap \left\langle G_3 \right\rangle )) = \dim(\left\langle G_1 \right\rangle \cap \left\langle G_2 \right\rangle \cap \left\langle G_3 \right\rangle) = 0 .
  \end{equation*}
\end{IEEEproof}

With Theorem~\ref{thm:fullrankA} we now have an explicit construction for the considered case by choosing the locators
$\alpha_i$ from a multiplicative group of size~$\gamma$. It remains to be shown that a field with a multiplicative group
of that order exists.

\begin{lemma}
  For any~$\gamma$ there exists a field~$\field_q$ such that there is a choice of locators for which the matrix~$A$ as in
  Theorem~\ref{thm:fullrankA} is of full rank.
\end{lemma}
\begin{IEEEproof}
  By Theorem~\ref{thm:fullrankA} the matrix~$A$ is always of full rank if the locators are chosen to be of order~$\gamma$.
  A field~$\field_q$ contains a multiplicative group of order~$\gamma$ if~$\gamma | q-1$, where~$q$ is a power
  of a prime. By Dirichlet's theorem \cite{Dirichlet1837} there are infinitely many primes of the
  form~$p = l+m \gamma$, if~$\gamma$ and~$l$ are coprime. For~$l=1$ any~$\gamma$ is coprime and it follows that for
  any~$\gamma$ there are infinitely many primes~$p$ such that~$\gamma | p-1$.
\end{IEEEproof}

For applications in data storage, the most interesting fields to consider are those of characteristic~$2$. Whether a
field~$\field_{2^s}$ for which a construction as described in Theorem~\ref{thm:fullrankA} is possible exists, depends on
the existence of a \textit{Mersenne number}~$M_p = 2^p-1$ such that~$\gamma | M_p$. As all Mersenne numbers are odd, so
are all their divisors and it follows that the construction over a field~$\field_{2^s}$ is only feasible for
odd~$\gamma$, \emph{i.e.}, for~$4 \nmid k$.  
\begin{table}
  \centering
    \caption{Results of computer search for different parameters. The column $P_{\mathrm{full}}$ gives the probability of the rank of $A$ being full obtained from checking $10000$ random choices of locators from the respective field.}
  \begin{tabular}{CCCCC}
    k&M=\epsilon & N & q & P_{\mathrm{full}}\\ \hline
    2&1&3&16&>0.99\\
    2&1&3&64&>0.99\\
    2&1&3&256&>0.99\\
    4&1&3&16&0.93\\
    4&1&3&64&0.98\\
    4&1&3&256&>0.99\\
    8&1&3&64&0.95\\
    8&1&3&256&>0.99\\
    16&1&3&64&0.98\\
    16&1&3&256&>0.99\\
    3&2&5&16&0.69\\
    3&2&5&64&0.89\\
    3&2&5&256&>0.99\\
    6&2&5&64&0.98\\
    6&2&5&256&>0.99\\
    9&2&5&64&0.97\\
    9&2&5&256&>0.99\\
    18&2&5&256&>0.99\\
    4&3&7&64&0.98\\
    4&3&7&256&>0.99\\
    8&3&7&64&0.98\\
    8&3&7&256&>0.99\\
    16&3&7&256&>0.99
  \end{tabular}
  \label{tab:fullranksim}
  \vspace{-20pt}
\end{table}

For certain parameters Theorem~\ref{thm:fullrankA} gives an explicit choice of locators such that the matrix~$A$ as defined in Definition~\ref{def:Amat} is of full rank. In general it is an open problem whether matrices of such structure are of full rank, however as shown in Table~\ref{tab:fullranksim}, computer searches suggest that the probability is high for a random choice of locators from a sufficiently large field.

\end{document}